\documentclass[journal]{IEEEtran}

\usepackage{courier}
\usepackage{amsmath}
\usepackage{amssymb}
\usepackage{wasysym}
  
\usepackage[]{graphicx,color}
\definecolor{red}{rgb}{0,0,0}
\definecolor{blue}{rgb}{0,0,0}

\usepackage[caption=false,font=footnotesize]{subfig}
\usepackage[ruled,vlined]{algorithm2e}
\usepackage{boxedminipage}
\usepackage{multirow}
\usepackage{wasysym}
\usepackage{enumerate}

\newtheorem{theorem}{Theorem}

\newtheorem{proposition}{Proposition}

\newtheorem{result}{Result}

\begin{document}
\title{Offering Supplementary Network Technologies: \\ Adoption Behavior and Offloading Benefits}
\author{Carlee Joe-Wong, Soumya Sen, and Sangtae Ha
\thanks{C. Joe-Wong is with the Program in Applied and Computational Mathematics, Princeton University, Princeton, NJ, USA. S. Sen is with the Carlson School of Management, University of Minnesota, Minneapolis, MN, USA. S. Ha is with the Interdisciplinary Telecom Program and Department of Computer Science, University of Colorado, Boulder, CO, USA. A preliminary version of this work appeared at IEEE INFOCOM 2013.}}

\maketitle

\begin{abstract}
% The rapid growth in data demand has led to significant congestion on wireless networks, forcing ISPs to introduce new technologies and pricing tools to maintain their economic viability.  In particular, offloading traffic onto a supplementary, better quality network technology, e.g., offloading from 3G or 4G to WiFi and femtocells, have emerged as mechanisms for congestion alleviation.
% In many industries, one technology may supplement another by offering enhanced features or quality. This base and supplemental technology relationship is increasingly evident in the mobile data market:
To alleviate the congestion caused by rapid growth in demand for mobile data, {\color{red} wireless service providers (WSPs) have begun encouraging users to offload some of their traffic onto supplementary network technologies}, e.g., offloading from 3G or 4G to WiFi or femtocells.  With the growing popularity of such offerings, a deeper understanding of the underlying economic principles and their impact on technology adoption is necessary.  To this end, we develop a model for user adoption of a base technology {\color{blue} (e.g., 3G)} and a bundle of the base plus a supplementary technology {\color{blue} (e.g., 3G + WiFi).  Users individually make their adoption decisions based on several factors, including the technologies' intrinsic qualities}, {\color{red} negative} congestion externalities from other subscribers, and the flat access rates that a {\color{red} WSP} charges.  {\color{blue} We then show how these user-level decisions translate into aggregate adoption dynamics and prove that these converge to a unique equilibrium for a given set of exogenously determined system parameters.}  We fully characterize these equilibria and {\color{blue} study} adoption behaviors of interest to a {\color{red} WSP}.  We then derive analytical expressions for the revenue-maximizing prices and optimal coverage factor for the supplementary technology and examine some {\color{red} resulting non-intuitive user adoption behaviors}.  Finally, we {\color{blue} develop a mobile app to collect} empirical 3G/WiFi usage data and numerically investigate the profit-maximizing adoption levels when a {\color{red} WSP} accounts for its cost of deploying the supplemental technology and savings from offloading traffic onto this technology.  

%Recently, a rapid increase in demand has led to significant congestion on mobile data networks.  While many solutions to this problem have been proposed and implemented, none have completely solved it.  In this work, we investigate the viability of offloading mobile data traffic onto a supplementary, better quality network technology, e.g., offloading 4G traffic to split-spectrum femtocells.  We suppose that ISPs charge flat monthly access rates for each of the two technologies and study the adoption patterns of heterogeneous users.  We find that user adoption levels converge to a unique equilibrium point as a function of the user preferences, the access prices charged, and the area covered by the supplemental technology.  In particular, as this technology's coverage area increases, its adoption may decrease due to increased congestion on its network.  Given this non-intuitive behavior, we then investigate an ISP's profit-maximizing prices and the optimal coverage area of the supplemental technology.  We find that the ISP maximizes its revenue by offering full coverage of the supplemental technology.  To account for ISP costs, we collect 3G and WiFi usage data from twenty users and estimate the benefits of offloading traffic to the supplemental technology.  We numerically characterize the profit-maximizing adoption levels when the ISP accounts for these savings from offloading and the cost of deployment for the supplemental technology.
\end{abstract}

%\begin{equation}
%\frac{\left\langle\left(x_1^{-\alpha}, x_{1 + 2}^{-\alpha},\ldots, x_n^{-\alpha}\right), \eta\right\rangle}{\left\langle\left(x_1^{-\alpha}, x_{1 + 2}^{-\alpha},\ldots, x_n^{-\alpha}\right), \frac{\left(1, 1,\ldots, 1\right)}{\sqrt{n}}\right\rangle}
%\end{equation}

\section{Introduction}\label{sec:intro}

%\subsection{Context and Motivation}

Successful technologies are often followed by other supplemental technologies, which when combined with the original enhance its features and quality. {\color{red} In the context of networks, one could think of 3G or 4G traffic being offloaded to {\color{blue} WiFi or femtocells respectively}. Yet adoption of these supplemental and base technologies depends not just on their access prices, but also on the externalities that users of each technology impose on others.}
% Upon introducing an additional access cost for users of the supplemental technology, one can then study the adoption of the supplemental and base technologies in a generic economic framework. Yet considering network access technologies also requires accounting for network externalities.
For example, as more users adopt a {\color{blue} base} technology like 3G or 4G, congestion on the network increases, which can in turn reduce users' utility from using that technology.

The presence of negative network externalities affords network operators an opportunity to improve services on their base technology by offloading users onto a supplemental technology. Indeed, {\color{red} Wireless Service Providers (WSP)s are already beginning to do so}: for instance, in the United States, Verizon has begun to offer femtocells in order to supplement its 4G network capacity \cite{verizonfemto}, while AT\&T has deployed WiFi hotspots in New York to manage persistent 3G congestion \cite{attwifi}. In light of rapid growth in the projected demand for mobile data \cite{cisco}, {\color{red} WSPs} are likely to continue using such supplemental networks to curb network congestion \cite{sen2012incentivizing}.  Indeed, as mobile demand keeps growing, {\color{blue} WSPs are beginning} to charge consumers for access to these supplemental networks. % Already, T-Mobile and Virgin Mobile respectively charge their subscribers an extra \$20 and \$15 a month for WiFi access \cite{wifialliance,TMobile,virgin}. {\color{red} Smaller carriers like Republic Wireless in the U.S. are beginning to explicitly use WiFi for most data transfers and calls, falling back on cellular connectivity only when WiFi is unavailable \cite{wifi_gigaom}.}
{\color{blue} For instance, Orange offers a \pounds2 bundle to some of their customers for access to WiFi hotspots \cite{orange_wifi}.} Given these developments, there is a need for economic models that can help determine \emph{how to price} access to such base and supplementary technologies and the implications of those pricing decisions.  

% Yet usage-based pricing is not enough--data traffic tends to concentrate at certain peak hours of the day, while at other times the network is underutilized.  To fully mitigate congestion, {\color{red} WSP}s must shift traffic in the peak hour, either to other times of the day or to other networks \cite{sigcomm}.
% In this work, we develop a generic model to study such schemes for \emph{offloading} traffic from one technology to a supplemental, better quality one, which nevertheless may not be always available.

% Our models apply to the offloading of 3G traffic to WiFi, or a split (but not dedicated) spectrum offloading of 4G traffic to femtocells.

% \subsection{{\color{red} WSP} Scenarios and User Adoption}
%In our adoption model, users may adopt the base technology, no technology, or a bundle of both technologies.  We assume that they have heterogeneous valuations of each technology and take into account congestion effects as technology adoption increases.  %Within this setting, user adoption over time converges to a steady-state equilibrium,  which the network quality is balanced by access prices and congestion from other users.

%\subsection{Results and Example Scenarios}

Our work is inspired by two research areas: the study of user technology adoption and that of network offloading {\color{red} and pricing}.  Though both areas have separately received considerable attention from economics and networking researchers, our contribution lies in incorporating user{\color{blue} -level} adoption models to study technology subscription dynamics and the {\color{red} consequent revenue and cost} tradeoffs {\color{blue} of offering} a supplementary technology.
We seek to understand user adoption decisions between a generic base technology and a bundled offering of a base plus supplemental technology; users may adopt the base technology, no technology, or the bundle of both technologies.
% {\color{red} WSP}s' goal in offering WiFi hotspots or femtocells is twofold: by charging extra for this service, they may increase their revenue, while also benefiting from traffic offloading.  Yet offering such networks comes at a cost, which may not be completely offset by the {\color{red} WSP} benefits.  And in fact such hotspots are quite localized; they do not offer coverage in all locations, which limits the potential for offloading traffic.  Our work introduces a mathematical framework in which to understand these various effects and the corresponding adoption behavior. In particular, we consider the following scenarios: {\color{red} describe all of the results}
In particular, we explore the following example scenarios:
{\color{blue} \begin{IEEEitemize}
\item
\emph{Will increasing the coverage area of a supplemental network always lead to more users adopting it?} Suppose that a {\color{red} WSP} wishes to expand its supplemental network to offload more traffic from the base technology, but cannot change its pricing structure due to exogenous factors, e.g., the presence of a major competitor.  We derive conditions under which increased supplemental network coverage will \emph{decrease} its adoption: at the new equilibrium, each user offloads more traffic, which can increase overall congestion on the supplemental network and induce some users to drop the bundled service (Result \ref{prop:x2eta}). We show that this decrease may occur even when the {\color{red} WSP} offers revenue-maximizing prices (Result \ref{prop:optx}).
\item
\emph{Can increasing the base technology's access price reduce its adoption?} Consider a {\color{red} WSP} trying to induce some users to leave its base technology's network by increasing the access price.  In some scenarios, this move can increase the base technology's adoption: increasing the price of base and the base + supplementary bundle by the same amount can lead some users dropping from their bundled subscriptions to only the base technology (Result \ref{prop:p3G}).
%\item
%An {\color{red} WSP} expands its femtocell network while offering revenue-maximizing 4G and femtocell access prices.  Yet as femtocell coverage increases, we identify conditions under which femtocell adoption decreases due to increased congestion.  However, the revenue-maximizing femtocell price also increases in this scenario, leading both 4G adoption and {\color{red} WSP} revenue to increase.
\end{IEEEitemize}
Given these adoption behaviors, we then consider a {\color{red} WSP}'s optimal operating point. In our framework, the {\color{red} WSP} may influence user adoption with three variables: the access prices of the two technologies, and the coverage area of the supplementary technology (e.g., supplementary WiFi access may not be available at all locations). We first consider a generic model of {\color{red} WSP} revenue and derive analytical expressions for the revenue-maximizing prices and coverage of the supplemental technology. We then focus on the scenario of offloading traffic from base to supplementary technology, estimating the offloading benefits and deployment costs with empirical usage data. We consider the following questions:
\begin{IEEEitemize}
\item
\emph{What is the optimal coverage area of a supplemental network?} Suppose that a {\color{red} WSP} seeks to maximize its revenue by optimizing the supplemental technology's coverage area and the base and bundle access prices.  We show that revenue is maximized with full supplemental network coverage (i.e., available everywhere) (Prop. \ref{prop:rev_partial}).
\item
\emph{Will users always adopt both the base and bundled technologies at the revenue-maximizing prices?} Suppose that the WSP cannot expand its supplemental network coverage, e.g., due to investment costs. We show that the base and bundle offers will both have users unless the supplementary network coverage area is sufficiently small and the base technology experiences sufficiently severe congestion externalities (Result \ref{prop:revenuemax}).
\item
\emph{Will increasing the supplemental technology's coverage area always increase the total adoption of the base and bundled technologies?}
Suppose that a WSP increases its supplementary network coverage in an attempt to increase its revenue. We derive necessary and sufficient conditions under which the total adoption will decrease as this coverage increases (Result \ref{prop:optx}).
\item
\emph{Will deployment costs affect the optimal coverage and resulting adoption of the base plus supplementary bundle?} While increasing the deployment costs of the supplementary technology decreases its optimal coverage area, it also decreases the potential congestion on this supplementary network. Consequently we find that the adoption of the supplementary technology can increase with its deployment costs, even when choosing profit-maximizing prices and coverage area (Fig. \ref{fig:costdeploy}).
\end{IEEEitemize}
}
To answer these questions, we develop an analytical framework that incorporates individual users' adoption decisions, instead of {\color{blue} simply} considering the aggregate adoption dynamics. {\color{blue} U}sers have heterogeneous valuations of each technology's quality and account for the negative externalities of congestion effects as a technology's adoption increases. In lieu of modeling details specific to a given technology, we utilize a generic externality model and investigate possible adoption behaviors; we also show that our qualitative results hold for some more specific externality models.  {\color{blue} We consider} a monopolistic WSP, reflecting the near-monopolistic conditions in many wireless Internet markets. For instance, in 2012 the dominant U.S. carriers AT\&T and Verizon reported near-negligible churn rates \cite{wsj_churn}, indicating that these providers essentially function as monopolists for their respective customer bases. {\color{blue} Instead of considering competition between such WSP, we consider different data plans that each WSP can offer. In Appendix \ref{sec:multiple}, we show an extension to multiple competing WSPs.}
In Section \ref{sec:related}, we compare our work to similar contributions in the literature, focusing on network adoption and offloading works.  We then introduce and analyze our model in Section \ref{sec:usage}, and examine some interesting adoption outcomes in Section \ref{sec:guidelines}.  Finally, in Section \ref{sec:profit}, we consider users' equilibrium adoption levels when a {\color{red} WSP} maximizes its revenue and profit.
We conclude the paper in Section \ref{sec:conclusion}.  Proofs of all propositions, theorems and results {\color{red} are available in Appendix \ref{sec:proofs}}.

% We assume that the number of users at any particular point in the area under consideration is a function of the distance from the nearest WiFi base station; there are more users closer to these base stations.

% Users derive utility from both 3G and WiFi coverage as a function of the price charged and the number of users using that technology.  Some heterogeneity is included, so that users derive different intrinsic utilities from the two technologies; we assume a uniform distribution of this parameter in the utility function.

\section{Related Work}\label{sec:related}

Our work relates to two topics in the literature: technology adoption dynamics and the economics of offloading traffic to supplemental networks (e.g., 3G/4G to WiFi/femtocells). % In this section, we give an overview of relevant prior works in these two areas, and contrast their analyses with distinguishing features of our adoption model and analysis of {\color{red} WSP} profit.

% Negative versus positive externalities

\subsection{Technology Adoption}

Many works in economics have studied technology adoption in various contexts \cite{geroski2000models,mahajan2000new}. %, or the survey in \cite{geroski2000models}.
Katz and Shapiro, for instance, consider competing network technologies with positive externalities in a homogeneous user population \cite{katz1986technology}, while Cabral \cite{cabral1990adoption} presents a diffusion model for a single technology{\color{red} 's} adoption by users with heterogeneous network valuations.  Economides and Viard \cite{economides2007pricing} provide a static analysis for the adoption of two complementary technologies with positive externalities and heterogeneity in user evaluations.  In the context of networks, Joseph et al. \cite{joseph2007modeling} study the adoption of new network architectures, characterizing possible adoption trajectories for a continuum of identical utility-maximizing users. {\color{red} Jin et al. \cite{jin2008dynamics} and} Sen et. al. \cite{sen2010modeling} model heterogeneous users to study the dynamics of competition between two generic network technologies with positive network externalities, {\color{red} considering in particular the effects of converters \cite{sen2010modeling}}.   

While our work follows these in modeling user heterogeneity, it differs in that (a) we explicitly model the negative externality of network congestion, as opposed to the dominant positive externality generally considered in technology adoption models; (b) we consider a non-competitive scenario in which the supplementary technology is {\color{red} bundled with the base technology}{\color{blue} ; and (c) we optimize the system decision variables and investigate the resulting equilibrium adoption outcomes}.  % We investigate the impact of profit-maximizing pricing and coverage decisions on equilibrium adoption outcomes.

In another work related to ours, Ren, Park, and van der Schaar consider market entry and spectrum sharing decisions of femtocell providers \cite{ren2012entry}.  We do not focus on the entrant-incumbent interaction of two providers sharing a market; instead, we model the adoption of a supplementary technology offered by a monopolist provider and the resulting tradeoffs between deployment costs and savings from offloading. % Our work is thus focused on traffic offloading, unlike \cite{ren2012entry}.

\subsection{Traffic Offloading}

Shetty, Parekh and Walrand analyze user adoption of split- and common-spectrum 4G and femtocell networks to study {\color{red} WSP} revenue maximization \cite{shetty2009economics}.  They consider the utility of heterogeneous users under both spectrum sharing schemes and account for congestion effects with detailed throughput models.  However, \cite{shetty2009economics} does not consider {\color{red} WSP} costs or savings from offloading and relies on numerical studies due to the complexity of their throughput models. In our more generic framework, we take the spectrum sharing scheme as an input and consider the resulting adoption dynamics.

%; moreover, due to the complexity of user throughput models, their results are primarily numerical.

Other works have also studied traffic offloading, but without developing an analytical model of user adoption decisions.  For instance, \cite{yaiparoj2008economics} considers the problem of offloading 3G traffic to WiFi networks, focusing on the implications for {\color{red} WSP} revenue.  User adoption is here modeled using given demand functions, which depend on the prices of 3G and WiFi.  Offloading onto femtocell networks is studied in \cite{yun2012economic}, which considers {\color{red} WSP} revenue and social welfare under flat and usage-based pricing of both open and closed femtocell networks.  Our work contributes to these efforts by providing a generic analytical framework, complemented with data collected from real users, to study the role of economic and technological decisions on the possible outcomes of the adoption process.    

%A more empirical study of offloading's benefits is provided in \cite{wireless2020}, which studies optimal WiFi coverage for the New York and San Diego markets.  The authors summarize a business analysis of WiFi offloading with relevant cost and usage parameters, though without considering user choice.  They find that the optimal amounts of WiFi coverage differ for New York and San Diego; similarly, our work shows that market population density can affect adoption behaviors.

% Our model of user adoption is not unique; several papers, e.g. \cite{sen2010modeling, shetty2009economics}, have used a variation.  While some investigate an offloading scenario, such works generally assume \emph{homogeneous} users, while our model accounts for user heterogeneity.  Ours is also the first work to consider the scenario of 3G and WiFi networks and to develop comprehensive guidelines for {\color{red} WSP}s facing such a scenario.  To make these as realistic as possible, we use real data...{\color{red} these sentences may work better in the introduction}

\section{Technology Adoption Model}\label{sec:usage}

In this section, we introduce an analytic framework to model the dynamics of user adoption based on the user's utility of subscribing to the base and supplemental technologies, denoted as Technologies 1 and 2, respectively.
% \footnote{We choose this notation to emphasize that our model is \emph{generic}, and may be applied to any market with base and supplementary technologies.}
We consider a monopolist {\color{blue} WSP.  Users may choose to adopt the base technology (Technology 1), a bundle of the base and supplemental technologies (Technologies (1+2)), or neither technology; Technology 2 is never offered without Technology 1.} This choice is governed by the {\color{blue} utility} that each of the above options provides to the user, as derived in Section \ref{sec:utility}.
Users' subscription choices evolve over time (e.g., weeks or months) in response to changes in the network adoption and congestion levels. {\color{red} To reflect the timescale of these choices, changes in the adoption dynamics are based on users' overall quality of experience in using each network. In Section \ref{sec:utility}, we formulate these subscription dynamics, and in} Section \ref{sec:stable}, we show that exactly one asymptotically stable equilibrium exists for any given set of exogenous system parameters.
%either adopt neither technology, adopt only technology 1, or adopt both technologies (e.g., adding a WiFi subscription to their 3G data plan).  
 %We assume that users are myopic, making decisions based on the (known) throughput for each technology at any given time. 
%Section \ref{sec:throughput} then characterizes throughput degradation as a function of user adoption levels and finds analytical expressions for steady-state equilibrium points.  

\subsection{{\color{red} User Adoption Decisions}}\label{sec:utility}
% Utility Functions

%A user's subscription decision depends on the ``value" offered by the different wireless technologies.  This value depends on several factors, such as the intrinsic quality of the technology (e.g. the user's monetary valuation of the maximum throughput), the negative externality of congestion (i.e., reduced throughput), and the access price charged by the service provider.    
 
A user's value or utility from subscribing to a particular wireless technology depends on several factors, such as the intrinsic quality of the technology (e.g., the user's valuation of the maximum throughput), the negative externality of congestion (i.e., reduced throughput), and the access price charged by the service provider. Following \cite{sen2010modeling} and \cite{shetty2009economics}, we account for these factors in defining the utility functions 
associated with each technology adoption option. For the two options, the base and the bundle (base plus supplementary) plans, the respective utility functions are given by (\ref{eq:utility3g}) and (\ref{eq:utilitywf}); the utility of non-adoption is assumed to be zero. 
\begin{align}
U_1(t) = & \theta q_1 + T_1\left(x_1 + (1 - \eta)x_{1 + 2}\right) - p \label{eq:utility3g} \\
U_{1 + 2}(t) = & (1-\eta)\big(\theta q_1 + T_1\left(x_1 + (1 - \eta) x_{1 + 2}\right)\big) \nonumber \\
& {} + \eta\big(\theta q_2 + T_2\left(\eta x_{1 + 2}\right) \big) - \left(p + \Delta\right). \label{eq:utilitywf}
\end{align}

The above utility functions have three separate value components, as we discuss here. All are normalized to monetary units.
The intrinsic qualities of Technologies 1 and 2 are denoted by $q_i$, $i = 1,2$; we assume that the supplemental technology has a higher intrinsic quality than the base ($q_2 > q_1$).  For example, WiFi typically delivers much higher maximum throughput than 3G. While the actual throughput, delay, etc. at a given time can vary depending on the distance from a cell tower, obstacles, etc., {\color{blue} recall} that users make adoption decisions based on the \emph{overall} quality experienced over a subscription period. The valuation of this intrinsic quality is weighted by a random variable $\theta\in[0,1]$ to account for user heterogeneity.% {\color{red} We use the argument $v$, denoting the user's volume of data usage, to emphasize the dependence of the user's experienced quality on the amount of data they consume.} Users who stream a lot of video, for instance, {\color{red} will} have a higher $\theta$ value {\color{red} since video quality is sensitive to throughput}, while those who mainly surf the web will have a low $\theta$ value.
\footnote{The exact values of the $q_i$ parameters depend on the particular technology being considered, while the distribution of $\theta$ values can be estimated from established techniques in marketing research, e.g., conjoint analysis \cite{green1978conjoint}.} % {\color{red} To simplify notation, we omit the argument $v$ in the remainder of the paper, as it is exogenously determined and affects only the variable $\theta$.}
 
%We use $q_i$, $i = 1,2$, to denote the intrinsic qualities of technologies 1 and 2, and assume that the supplemental technology has a higher intrinsic quality than the base ($q_2 > q_1$).  To account for heterogeneity in user valuations of these qualities, we introduce a random variable $\theta\in[0,1]$; $\theta q_i$ then represents a given user's valuation of technology $i$.  Users who stream a lot of video, for instance, might have a high $\theta$ value, while those who mainly surf the web would have a low $\theta$ value.

{\color{blue} The supplemental technology, Technology 2, can have} a limited coverage area (e.g., users are not always within range of a hotspot or a femtocell) that determines the ``coverage factor'' $\eta$, such that {\color{blue} on average,} a fraction $\eta$ of {\color{blue} traffic from adopters of the technology bundle is offloaded to Technology 2's network. % As $\eta$ increases, each adopter of the bundle can thus offload more traffic onto Technology 2, which can increase its congestion.
We assume that users are distributed uniformly throughout the coverage area, so that at any given location, the expected amount of traffic offloaded over time equals $\eta$, multiplied by the total traffic of the bundle adopters. Assuming that the distribution of usage volume for these adopters is independent of their valuation types $\theta$, the total usage of the bundled adopters is then proportional to the number of such adopters. Moreover, since users are distributed uniformly throughout the coverage area, each user experiences the same amount of congestion from others. As is generally the case with supplementary technologies such as WiFi, we suppose that adequate infrastructure (e.g., device antennas) and handoff protocols are in place to allow users connectivity to the two technologies with no cost of switching between them. Nearly all mobile phones and tablets, for instance, can automatically switch from cellular to WiFi networks.% % --as discussed below, the presence of more users implies that a given user will receive a smaller share of the technology's resource.
%\footnote{While the congestion experienced by a given user at a given time and location will likely depend %on the number of other users as well as their usage vloume, the long-term time-averaged congestion %experienced by a user under homogeneous mobility assumption is simply proportional to the number of %other interferring users, $\eta x_{1+2}$.}. 
 % Thus, users who adopt the bundle are assumed to switch freely between the two technologies, subject to standard admission controls. For instance, adopters of the bundle could only make use of Technology 2's network if they had adequate coverage at their location and the network was not excessively congested. While users might experience a small inconvenience at the time of switching, we do not explicitly model this effect as it has relatively little impact on their overall utility from the two technologies. % \footnote{We assume that users are uniformly distributed around the coverage area and homogeneous in their usage volume.  Thus, the amount of traffic offloaded to each of Technology 2's hotspots is given by $\eta x_{1 + 2}$.}  %For instance, users might send only half of their traffic over WiFi due to a lack of WiFi access during the day.
%

Given the above assumptions, the average amount of traffic on Technology 2 is proportional to} the fraction of users adopting the technology bundle, multiplied by $\eta$.  We let $x_1(t)$ denote the fraction of users adopting only Technology 1 at time $t$ and $x_{1 + 2}(t)$ the fraction of users adopting the bundle.\footnote{We note that $x_1(t)$, $x_{1 + 2}(t)$, and $x_1(t) + x_{1 + 2}(t)\in[0,1]$.}  Thus, the {\color{blue} volume of traffic} on Technology 2 is $\eta x_{1 + 2}(t)$. The {\color{blue} amount of traffic on Technology 1 from users adopting only Technology 1 is} $x_1(t)$, while the  {\color{red} amount from bundle adopters} is $(1 - \eta)x_{1 + 2}(t)$. {\color{red} The total traffic volume on Technology 1 is then} $x_1(t) + (1 - \eta)x_{1 + 2}(t)$.
We use decreasing functions $T_1\big(x_1(t) + (1 - \eta) x_{1 + 2}(t)\big)$ and $T_2\big(\eta x_{1 + 2}(t)\big)$ to represent the {\color{red} negative externality} as a function of the {\color{red} number of users} for Technologies 1 and 2 respectively.\footnote{\color{blue} While users can also experience positive externalities, in the context of mobile data this positive externality is dominated by the content that users can access. Since data plans, unlike in-network voice calls, rarely offer extra benefits for communicating with other users on the same technology, the positive externality can be modeled as a constant term encapsulated in the similarly constant access price of the base technology. We suppose that plans for voice calls or text messages are orthogonal to a user's data plan.}
% we suppose that the negative externality dominates the positive one. i.e., that $T_1(x) > R_1(x)$, $T_2(x) > R_2(x)$, and that $T_1(x) - R_1(x)$ and $T_2(x) - R_2(x)$ are decreasing functions for all $x$. Without loss of generality, in the remainder of the paper we therefore omit the $R_1$ and $R_2$ externality functions, replacing the negative externalities $T_1$ and $T_2$ by $T_1 - R_1$ and $T_2 - R_2$ respectively.
%
We can interpret these negative externalities as the decrease in user utility due to lower throughput from {\color{blue} higher usage volumes.} % multiple users on the technology. As the number of users increases, the base station or access point needs to schedule users so as to fairly share the link capacity; sharing the technology's capacity with more users therefore decreases a given user's throughput and increases the delay of data transfers. Moreover, most networks can serve only a limited number of users, so users may find it more difficult to consistently establish a network connection as the number of other users increases.}\footnote{We limit our model to scenarios in which {\color{red} Technology 1 users do not affect Technology 2's throughput}, e.g., WiFi speeds are not affected by 3G traffic.}
% The quantity of traffic on Technology 1, $x_1 + (1 - \eta)x_{1 + 2}$, reflects the number of users adopting only Technology 1 ($x_1$), plus the number of users who adopt the bundled Technologies (1 + 2), multiplied by one minus the coverage factor ($(1 - \eta)x_{1 + 2}$). Similarly, the quantity of traffic on Technology 2 is simply the coverage factor, multiplied by the number of adopters of the bundled technology ($\eta x_{1 + 2}$).

The wireless service provider prices the access for the two options at $p$ for the base technology and $p+\Delta$ for the base plus supplemental technology bundle (i.e., $\Delta$ is the extra price that a user pays for the bundled option). For ease of notation, the time arguments of $x_1(t)$ and $x_{1 + 2}(t)$ will be assumed from here on to be implicit in the utility functions (\ref{eq:utility3g}) and (\ref{eq:utilitywf}).

%We add the quality, throughput and price terms described above to find the utility of adopting technology 1 and adopting both technologies:
%\begin{align}
%U_1(t) &= \theta q_1 + T_1\left(x_1 + (1 - \eta)x_{1 + 2}\right) - p \label{eq:utility3g} \\
%U_{1 + 2}(t) &= (1-\eta)\big(\theta q_1 + T_1\left(x_1 + (1 - \eta) x_{1 + 2}\right)\big) \nonumber \\ &+ \eta\big(\theta q_2 + T_2\left(\eta x_{1 + 2}\right)\big) - \Delta - p. \label{eq:utilitywf}
%\end{align}

% where $q_1$ and $q_2$ represent the intrinsic quality of 3G and WiFi respectively. For instance, these parameters could capture the variability of 3G service as compared to WiFi ($q_1 < q_2$), or the ``baseline'' utility that users give to 3G and WiFi service.  The functions $T_1$ and $T_2$ represent the congestion of 3G and WiFi coverage networks respectively.  Congestion depends on the fraction of total users subscribing to that technology ($x_1$ and $x_{1 + 2}$ for 3G only and 3G + WiFi respectively).  We assume that the positive externality of connectivity $C$ is essentially constant and included in the $q$ parameters, as 3G and WiFi both allow connectivity to the same number of users.

% The flat access prices for 3G and WiFi respectively are denoted $p$ and $\Delta$, with $\eta$ a parameter denoting the percentage of area covered by WiFi.  The parameters $\theta$ are user-specific and assumed uniformly distributed between 0 and 1.  These represent users' intrinsic preferences for the two technologies.

Given these functions, we can find the threshold value of $\theta$, $\theta_{(1,0)}$, for which users will prefer to adopt Technology 1 (i.e., $U_1 > 0$).  Similarly, we can also find the value of $\theta_{(1 + 2,1)}$ for which Technology 1 users will prefer the bundle of Technologies (1 + 2) (i.e., $U_{1+2} > U_1 > 0$).  Each $\theta$ threshold is a (time-dependent) function of $x_1$ and $x_{1 + 2}$.
The threshold $\theta_{(1,0)}$  for preferring Technology 1 occurs when $U_1 = 0$, i.e.,
%\begin{equation*}
% \theta_{(1,0)} q_1 + T_1(x_1) - p = 0
% \end{equation*}
% and
 \begin{equation}
 \theta_{(1,0)} = \frac{p - T_1(x_1 + (1 - \eta) x_{1 + 2})}{q_1}.
 \label{eq:theta3G}
 \end{equation}
 The threshold $\theta_{(1 + 2,1)}$ for adopting Technology 2 in addition to Technology 1 occurs when $U_{1+2} = U_1 \geq 0$, i.e.,
% \begin{equation*}
% \theta q_1 + (1-\gamma_1)\frac{-R_1\log(1-\gamma_1)}{\gamma_1} = \theta q_2 + (1-\gamma_2)\frac{-R_2\log(1-\gamma_2)}{\gamma_2} - \frac{\Delta}{\eta}
% \end{equation*}
% and
 \begin{equation}
 \theta_{(1 + 2,1)} {\color{red} =} \frac{T_2\left(\eta x_{1 + 2}\right) - T_1\left(x_1 + (1 - \eta) x_{1 + 2}\right) - \frac{\Delta}{\eta}}{q_1 - q_2}.
 \label{eq:theta3GWF}
 \end{equation}
 Finally, we solve for the threshold $\theta_{(1 + 2,0)}$ above which users will prefer to adopt both technologies, rather than have no connectivity.  This occurs when $U_{1 + 2} = 0$, or
 \begin{align}
 \theta_{(1 + 2,0)} = &\frac{-(1 - \eta)T_1\left(x_1 + (1 - \eta) x_{1 + 2}\right) - \eta T_2(\eta x_{1 + 2})}{(1 - \eta)q_1 + \eta q_2} \nonumber \\
 &+ \frac{p + \Delta}{(1 - \eta)q_1 + \eta q_2}.
 \label{eq:thetaWF}
 \end{align}
 
In the remainder of this paper, we follow existing literature {\color{red} \cite{jin2008dynamics,sen2010modeling,ren2012entry}} by taking the throughput degradation functions $T_1$ and $T_2$ to be linear, i.e., $T_i(x) = -\gamma_i x$, $i = 1,2$ for some positive constants $\gamma_i$. {\color{red} In practice, linear functions will only approximate the ``true,'' nonlinear throughput degradation functions, e.g., $T_i(x) = -x^\alpha$ for $\alpha \in (0,1)$. We use this approximation to keep our approach generic rather than technology-specific, as the exact algorithm for distributing resources among multiple users varies by the type of technology (i.e., 3G, 4G, WiFi, etc.) and sometimes the deployment vendor. In Appendix \ref{sec:throughput}, we derive analytical bounds on the approximation error for typical throughput functions and show that qualitatively similar results are achieved with nonlinear functions.} With the linear $T_1$ and $T_2$, (\ref{eq:theta3G}-\ref{eq:thetaWF}) become
\begin{align}
 \theta_{(1,0)} = &\frac{p + \gamma_1(x_1 + (1 - \eta) x_{1 + 2})}{q_1}.
 \label{eq:theta3Glin} \\
 \theta_{(1 + 2,1)} = &\frac{-\eta\gamma_2 x_{1 + 2} + \gamma_1\left(x_1 + (1 - \eta) x_{1 + 2}\right) - \frac{\Delta}{\eta}}{q_1 - q_2}.
 \label{eq:theta3GWFlin} \\
 \theta_{(1 + 2,0)} = &\frac{(1 - \eta)\gamma_1\left(x_1 + (1 - \eta) x_{1 + 2}\right) + \eta^2 \gamma_2x_{1 + 2}}{(1 - \eta)q_1 + \eta q_2} \nonumber \\
 &+ \frac{p + \Delta}{(1 - \eta)q_1 + \eta q_2}.
 \label{eq:thetaWFlin}
 \end{align}

For given adoption levels $x_1$ and $x_{1 + 2}$, the ordering of these threshold values (\ref{eq:theta3Glin}-\ref{eq:thetaWFlin}) determines whether a user of type $\theta$ is willing to adopt a particular technology.  Thus, we can determine the fraction of users $H_1\big(x_1(t), x_{1 + 2}(t)\big)$ and $H_{1 + 2}\big(x_1(t), x_{1 + 2}(t)\big)$ willing to adopt Technology 1 and Technologies (1 + 2) respectively.  In doing so, we recall that $\theta\in[0,1]$; for instance, if $\theta_{(1,0)} < 0$, \emph{all} users receive positive utility from adopting Technology 1. We thus let $\left[\cdot\right]_{[0,1]}$ denote the projection onto the $[0,1]$ interval.\footnote{That is, $[y]_{[0,1]} = y$ if $y\in[0,1]$, 0 if $y < 0$, and 1 if $y > 1$.}

We first consider the case $\theta_{(1,0)} < \theta_{(1 + 2,0)}$, i.e., the threshold for preferring the base technology to no adoption is smaller than that of preferring both technologies to no adoption.  We show that $\theta_{(1 + 2, 1)} > \theta_{(1 + 2, 0)}$; thus, if a user receives positive utility from Technology 1 and increases it by adopting Technology 2 as well $\left(\theta_{(1,0)} < \theta_{(1 + 2, 1)} < \theta\right)$, she cannot receive negative utility from adopting both technologies $\left(\theta_{(1,0)} < \theta_{(1 + 2, 1)} < \theta < \theta_{(1 + 2, 0)}\right)$. {\color{blue} The threshold orderings must then satisfy the following:}

\vspace{0.04in}
 \begin{proposition}\label{prop:thetas}
If $\theta_{(1,0)} < \theta_{(1 + 2,0)}$, then $\theta_{(1 + 2,0)} < \theta_{(1 + 2,1)}$.  If $\theta_{(1 + 2,0)} < \theta_{(1,0)}$, then $\theta_{(1 + 2,1)} < \theta_{(1 + 2,0)}$.
\end{proposition}
\vspace{0.07in}

Thus, if $\theta_{(1,0)} < \theta_{(1 + 2,0)}$, the fraction of users $H_1$ willing to adopt Technology 1 equals the fraction for whom $\theta_{(1,0)} < \theta < \theta_{(1 + 2, 1)}$, and the fraction $H_{1 + 2}$ willing to adopt Technologies (1 + 2) equals those for whom $\theta_{(1 + 2, 1)} < \theta$.  {\color{blue} Assuming $\theta$ is uniformly} distributed between 0 and 1, we have
\begin{align}
H_1\big(x_1, x_{1 + 2}\big) &= \left[\theta_{(1 + 2, 1)}\right]_{[0,1]} - \left[\theta_{(1,0)}\right]_{[0,1]}, \nonumber \\
H_{1 + 2}\big(x_1, x_{1 + 2}\big) &= 1 - \left[\theta_{(1 + 2, 1)}\right]_{[0,1]}. \label{eq:H12}
\end{align}
If the thresholds are reversed, i.e., $\theta_{(1 + 2, 0)} < \theta_{(1,0)}$, then we may use Prop. \ref{prop:thetas} to derive
\begin{equation}
H_1\big(x_1, x_{1 + 2}\big) = 0,\, H_{1 + 2}\big(x_1, x_{1 + 2}\big) = 1 - \left[\theta_{(1 + 2, 0)}\right]_{[0,1]}.
\label{eq:H21}
\end{equation}
Following standard economic models, we assume a negligible cost of switching between the adoption choices \cite{sen2010modeling,ren2012entry}.
\begin{table}
\centering
 \renewcommand{\arraystretch}{1.2}
\caption{Expressions for $H_1$ and $H_{1 + 2}$ in different regions of $\left(x_1, x_{1 + 2}\right)$.}
\label{tab:regions}
\vspace{-0.1in}
\begin{tabular}{|c|c|c|c|}
\hline
& Conditions on $\theta$ & $H_1$ & $H_{1 + 2}$ \\ \hline
\multirow{2}{*}{a} & $\theta_{(1 + 2,0)} < \theta_{(1 + 2,1)} < 0$ &  \multirow{2}{*}{0} &  \multirow{2}{*}{1} \\
& $\theta_{(1 + 2,1)} < \theta_{(1 + 2,0)} < 0$ & & \\ \hline
b & $\theta_{(1,0)} < 0 < \theta_{(1 + 2,1)} < 1$ & $\theta_{(1 + 2,1)}$ & $1 - \theta_{(1 + 2,1)}$ \\ \hline
c & $0 < \theta_{(1,0)} < \theta_{(1 + 2,1)} < 1$ & $\theta_{(1 + 2,1)} - \theta_{(1,0)}$ & $1 - \theta_{(1 + 2,1)}$ \\ \hline
d & $0 < \theta_{(1,0)} < 1 < \theta_{(1 + 2,1)}$ & $1 - \theta_{(1,0)}$ & 0 \\ \hline
\multirow{2}{*}{e} & $0 < \theta_{(1 + 2,0)} < 1 < \theta_{(1,0)}$ & \multirow{2}{*}{0} & \multirow{2}{*}{$1 - \theta_{(1 + 2,0)}$} \\ 
& $0 < \theta_{(1 + 2,0)} < \theta_{(1,0)} < 1$ & & \\ \hline
f & $\theta_{(1,0)} < 0 < 1 < \theta_{(1 + 2,1)}$ & 1 & 0 \\ \hline
\multirow{2}{*}{g} & $1 < \theta_{(1,0)} < \theta_{(1 + 2,0)}$ & \multirow{2}{*}{0} & \multirow{2}{*}{0} \\
& $1 < \theta_{(1 + 2,0)} < \theta_{(1,0)}$ & & \\ \hline
\end{tabular}
\vspace{-0.15in}
\end{table}

\begin{figure}[t]
\centering
\includegraphics[width = 0.35\textwidth]{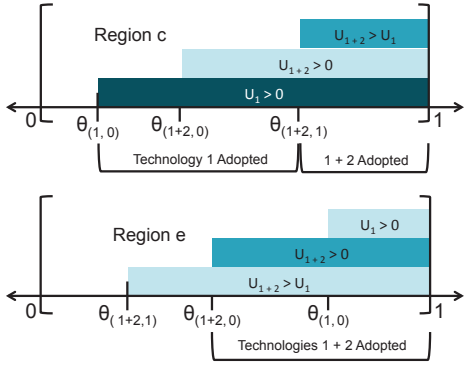}
\vspace{-0.05in}
\caption{Visualization of $\theta$ and $H$ values for regions c and e in Table \ref{tab:regions}.}
\label{fig:thetas}
\vspace{-0.15in}
\end{figure}

\begin{figure}[t]
\centering
\includegraphics[width = 0.4\textwidth]{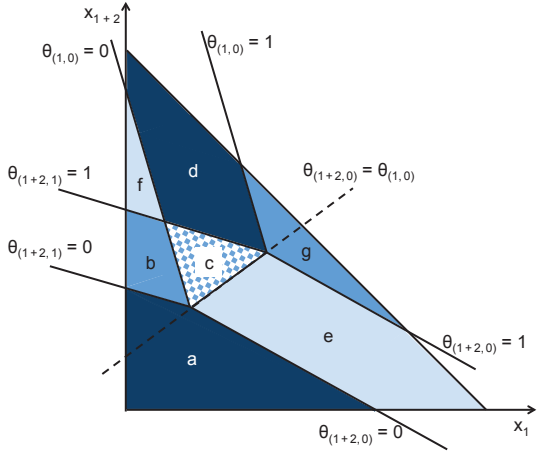}
\vspace{-0.1in}
\caption{Visualization of Table \ref{tab:regions}'s regions in terms of the adoption levels.}
\label{fig:regions}
\vspace{-0.25in}
\end{figure}

By dividing the dynamical space into seven regions, we can explicitly write out (\ref{eq:H12} - \ref{eq:H21}) as in Table \ref{tab:regions}.  Figure \ref{fig:thetas} visually represents the adoption expressions in two regions, and Fig. \ref{fig:regions} uses the $\theta$ threshold values (\ref{eq:theta3Glin} - \ref{eq:thetaWFlin}) to map them to the adoption levels $x_1$ and $x_{1 + 2}$.\footnote{A qualitatively similar figure with the same adjacent regions will be obtained even for nonlinear $T_1$ and $T_2$.}
The user dynamics are then
 \begin{align}
 \dot{x}_1(t) &= \rho\Big(H_1\big(x_1(t), x_{1 + 2}(t)\big) - x_1(t)\Big) \nonumber \\
 \dot{x}_{1 + 2}(t) &= \rho\Big(H_{1 + 2}\big(x_1(t), x_{1 + 2}(t)\big) - x_{1 + 2}(t)\Big),
 \label{eq:xdyn}
 \end{align}
where $\rho\in(0,1]$ {\color{blue} expresses the hazard rate, or the probability that a user who has not yet adopted a technology will do so at time $t$. If $\rho=1$, the fraction of users adopting each technology equals the fraction willing to adopt, less those who have already done so.}  Given these dynamics, we now derive the possible equilibrium points in each region, i.e., the values of $x_1$ and $x_{1 + 2}$ for which $H_1\big(x_1, x_{1 + 2}\big) = x_1$ and $H_{1 + 2}\big(x_1, x_{1 + 2}\big) = x_{1 + 2}$ for the $H$ expressions in Table \ref{tab:regions}.  Tables \ref{tab:equilibria} and \ref{tab:c} summarize the expressions for possible equilibria in each region{\color{blue} , along with constraints ensuring that the equilibria lie in their corresponding regions.}
 
 \begin{table*}
 \centering
 \renewcommand{\arraystretch}{1.4}
 \caption{Equilibrium points $\left(\overline{x}_1 \overline{x}_{1 + 2}\right)$ of the different regions in Table \ref{tab:regions}.}
 \label{tab:equilibria}
 \vspace{-0.1in}
 \begin{tabular}{|c|c|c|}
 \hline
 & $\left(\overline{x}_1, \overline{x}_{1 + 2} \right)$ & Region Constraints \\ \hline
 \multirow{2}{*}{a} &  \multirow{2}{*}{$(0,1)$} & $p + \Delta < -(1 - \eta)^2\gamma_1 - \eta^2\gamma_2$ \\
 & & $\Delta < \eta\left((1 - \eta) \gamma_1-\eta\gamma_2\right)$ \\ \hline
 
 \multirow{2}{*}{b} & \multirow{2}{*}{$\left(\frac{\eta\left((1 - \eta)\gamma_1 - \eta\gamma_2\right) - \Delta}{\eta\left(q_1 - q_2\right) - \eta^2\left(\gamma_1 + \gamma_2\right)}, \frac{\Delta - \eta\gamma_1 + \eta\left(q_1 - q_2\right)}{\eta\left(q_1 - q_2\right) - \eta^2\left(\gamma_1 + \gamma_2\right)}\right)$} & $\left(\eta\left(\gamma_1 + \gamma_2\right) - q_1 + q_2\right) p + \gamma_1\Delta < -\eta\gamma_1\gamma_2 + (1 - \eta)\gamma_1\left(q_1 - q_2\right)$ \\ 
 & & $\eta\left((1 - \eta)\gamma_1 - \eta\gamma_2\right) < \Delta < \eta\left(\gamma_1 + q_2 - q_1\right)$ \\ \hline
 
 c & See Table \ref{tab:c}. & See Table \ref{tab:c}. \\ \hline
 
 d & $\left(\frac{q_1 - p}{q_1 + \gamma_1}, 0\right)$ & $-\gamma_1 < p < q_1$, $\Delta + \frac{\eta\gamma_1p}{q_1 + \gamma_1} > \eta\left(q_2 - q_1 + \frac{\gamma_1q_1}{q_1 + \gamma_1}\right)$ \\ \hline
 
 \multirow{2}{*}{e} & \multirow{2}{*}{$\left(0, \frac{(1 - \eta)q_1 + \eta q_2 - p - \Delta}{(1 - \eta)q_1 + \eta q_2 + (1 - \eta)^2 \gamma_1 + \eta^2\gamma_2}\right)$} & $-(1 - \eta)^2\gamma_1 - \eta^2\gamma_2 < p + \Delta < (1 - \eta)q_1 + \eta q_2$\\ 
 & & $\eta\left(q_2 - q_1 - (1 - \eta)\gamma_1 + \eta\gamma_2\right)p - \left(q_1 + (1 - \eta)\gamma_1\right)\Delta > \eta^2 q_1\gamma_2 - \eta(1 - \eta)\gamma_1q_2$ \\ \hline
 % $p\left(q_1 - q_2 + D_e\right) + \Delta\left(\frac{q_1}{\eta} + D_e\right) > D_e\left((1 - \eta)q_1 + \eta q_2\right)$ \\ \hline
 
 f & $(1,0)$ & $p < -\gamma_1$, $\Delta > \eta \left(q_2 + \gamma_1 - q_1\right)$ \\ \hline
 
 \multirow{2}{*}{g} & \multirow{2}{*}{$(0,0)$} & $p > q_1$ \\
 & & $p + \Delta > (1- \eta)q_1 + \eta q_2$ \\ \hline
 
%  \multicolumn{3}{|l|}{$C_b = \eta\left(\gamma_1 + \gamma_2\right) - q_1 + q_2$,\;$C_e = \eta(q_2 - q_1) - \eta(1 - \eta)\gamma_1 + \eta^2\gamma_2$} \\ \hline
 % $D_e = \frac{(1 - \eta)q_2\gamma_1 - \eta q_1\gamma_2}{(1 - \eta)q_1 + \eta q_2 - (1 - \eta)^2\gamma_1 - \eta^2\gamma_2}$
 \end{tabular}
% \vspace{-0.05in}
 \end{table*}
 
 % We note that the equilibrium points and the conditions under which each equilibrium lies in its corresponding region are all linear in $p$ and $\Delta$.  Thus, no $p$ and $\Delta$ satisfy the region constraints for all regions.  For instance, the constraints $p < -\gamma_1 < 0$ in region f and $p > q_1 > 0$ in region g are incompatible.
 
 \begin{table*}
 \centering
 \renewcommand{\arraystretch}{1.2}
 \caption{Equilibrium points $\left(\overline{x}_1, \overline{x}_{1 + 2}\right)$ of region c in Table \ref{tab:regions}.}
 \label{tab:c}
 \vspace{-0.1in}
 \begin{tabular}{|c|c|}
 \hline
 \multirow{2}{*}{$\overline{x}_1$} & \multirow{2}{*}{$\frac{-\eta\gamma_2q_1 + (1 - \eta)\gamma_1q_2 + p\left(\eta\gamma_2 - (1 - \eta)\gamma_1 + q_2 - q_1\right) + \Delta\left(-(1 - \eta)\gamma_1 - q_1\right)/\eta}{-\gamma_1q_2 - \eta\gamma_1\gamma_2 + q_1\left((1 - \eta)\gamma_1 - \eta\gamma_2 + q_1 - q_2\right)}$} \\
 & \\ \hline
 \multirow{2}{*}{$\overline{x}_{1 + 2}$} & \multirow{2}{*}{$\frac{-\gamma_1q_2 + q_1\left(q_1 - q_2\right) + p\gamma_1 + \Delta\left(\gamma_1 + q_1\right)/\eta}{-\gamma_1q_2 - \eta\gamma_1\gamma_2 + q_1\left((1 - \eta)\gamma_1 - \eta\gamma_2 + q_1 - q_2\right)}$} \\
 & \\ \hline
 \multirow{3}{*}{Constraints} & $p\left(\eta\gamma_2 - (1 - \eta)\gamma_1 + q_2 - q_1\right) + \Delta\left(-(1 - \eta)\gamma_1 - q_1\right)/\eta < \eta\gamma_2q_1 - (1 - \eta)\gamma_1q_2$ \\
 & $p\gamma_1 + \Delta\left(\gamma_1 + q_1\right)/\eta < \gamma_1q_2 - q_1\left(q_1 - q_2\right)$ \\
 & $p\left(\eta\gamma_2 + \eta\gamma_1 + q_2 - q_1\right) + \Delta\gamma_1 > -\eta\gamma_1\gamma_2 + (1 - \eta)\gamma_1\left(q_1 - q_2\right)$ \\ \hline
 \end{tabular}
 \vspace{-0.15in}
 \end{table*}
 
 \subsection{Convergence and Stability}\label{sec:stable}
 
We now examine the stability of the equilibrium points in Tables \ref{tab:equilibria} and \ref{tab:c}:
\vspace{0.05in}
\begin{proposition}\label{prop:stable}
 Assuming that an equilibrium point exists, it is asymptotically stable.
 \end{proposition}
 \vspace{0.07in}
 
% Since all of our dynamics are linear, we can use the Jacobian of (\ref{eq:xdyn}) to assess the stability of each possible equilibrium point.  We assume that the conditions in Table \ref{tab:equilibria} are satisfied, so that the equilibrium exists and lies in its corresponding region.  Since the Jacobian matrix involves only the fixed $q$, $\eta$, and $\gamma$ constants, the constraints and equilibrium values in Table \ref{tab:equilibria} need not be considered.  Under these assumptions, we have the following proposition:

%In regions a, f and g, the Jacobian of the dynamical equation (\ref{eq:xdyn}) is the negative of the identity matrix.  Moreover, the boundary of each region, assuming linear throughput functions, is convex.  Thus, $x_1$ and $x_{1 + 2}$ converge linearly to the equilibrium point, yielding the following corollary to Prop. \ref{prop:stable}:
%\begin{corollary}\label{cor:stable}
%Assuming an equilibrium point exists in regions a, f, or g, any trajectory beginning in region a, f or g will stay in that region for all time and converge to the equilibrium point in Table \ref{tab:equilibria}.
%\end{corollary}
%
%We note that this method of proof cannot be applied to regions b, c, d or e, as the Jacobian in these regions has unequal eigenvalues.  Moreover, trajectories may begin in any region and converge to an equilibrium point in region a, f, or g.

We can use this result to show that for any set of exogenous parameters values and initial adoption levels, the adoption dynamics must converge to some stable equilibrium:
\vspace{0.05in}
\begin{proposition}\label{prop:period}
With the adoption dynamics (\ref{eq:xdyn}), no periodic orbit can exist: for any initial values $x_1(0)$ and $x_{1 + 2}(0)$, $x_1(t)$ and $x_{1 + 2}(t)$ converge to a stable equilibrium point.
\end{proposition}
\vspace{0.07in}

 While we assume that throughput degradation ($T_1$ and $T_2$) is linear in the previous section, the proofs of Props. \ref{prop:stable} and \ref{prop:period} depend respectively on only the Jacobian of the dynamics (\ref{eq:xdyn}) at given adoption levels $(x_1, x_{1 + 2})$.  Since our only assumption on the slopes $\gamma_1$ and $\gamma_2$ of the throughput degradation $T_1$ and $T_2$ is positivity, the Jacobian expressions at any given adoption levels are not affected by nonlinear forms of the $T_i$.  Thus, if $T_1$ and $T_2$ are continuously differentiable and strictly decreasing, both propositions still hold, i.e., for any initial values $x_1(0)$ and $x_{1 + 2}(0)$, $x_1(t)$ and $x_{1 + 2}(t)$ converge to a stable equilibrium point.
 
Moreover, only one such equilibrium point exists:
% One can also use the region constraints in Tables \ref{tab:equilibria} and \ref{tab:c} to show that only one equilibrium operating point may exist:
\vspace{0.05in}
\begin{theorem}\label{thm:unique}
For given values of the system parameters $q_1$, $q_2$, $\eta$, $\gamma_1$, $\gamma_2$, $p$, and $\Delta$, the adoption levels $x_1(t)$ and $x_{1 + 2}(t)$ converge to a unique, asymptotically stable equilibrium that does not depend on the initial values $x_1(0)$ and $x_{1 + 2}(0)$.
\end{theorem}
\vspace{0.07in}

In the remainder of the paper, we use $\overline{x}_1$ and $\overline{x}_{1 + 2}$ to denote the unique equilibrium adoption levels. While the \emph{overall} {\color{blue} adoption levels} $x_1$ and $x_{1 + 2}$ do not change at equilibrium, individual users may change their adoption decisions.

\section{Selected Adoption Behaviors}\label{sec:guidelines}

We now investigate the dependence of the equilibrium adoption on prices and the coverage factor.  Though the full equilibrium behaviors may be directly derived from Table \ref{tab:equilibria}, in this section we present some adoption outcomes and consequences that are of importance to a {\color{red} WSP}. Section \ref{sec:coverage} highlights conditions under which the total adoption can increase with {\color{blue} an increase in} the coverage factor, while Section \ref{sec:prices} derives conditions under which adoption of the base technology can increase with the base technology's access price. In {\color{red} Section \ref{sec:nonuniform_theta}, we show that similar behaviors occur when the user heterogeneity variable $\theta$ is non-uniformly distributed. Similar results are obtained with nonlinear throughput degradation functions $T_1$ and $T_2$, as shown in Appendix \ref{sec:throughput}.}

Our first observation is that full adoption $\left(\overline{x}_1 + \overline{x}_{1 + 2} = 1\right)$ {\color{blue} can only be achieved with subsidies ($p < 0$, $\Delta < 0$ or both)}, which may be seen by inspection of regions a, b, and f in Table \ref{tab:equilibria}. If some users' technology valuations $\theta q_i$ are sufficiently close to zero due to a small $\theta$ value, their utility functions (\ref{eq:utility3g}) and (\ref{eq:utilitywf}) will be negative unless the prices are negative (i.e., the {\color{red} WSP} offers adoption subsidies). Since we assume that $\theta$ is uniformly distributed, these subsidies are necessary for full adoption. In the remainder of this section, we consider only scenarios without full adoption $\left(\overline{x}_1 + \overline{x}_{1 + 2} < 1\right)$.

\subsection{Effect of the Coverage Factor}\label{sec:coverage}

 \begin{figure*}
 \centering
 \subfloat[Adoption as $\eta$ varies, $\overline{x}_1 > 0$ for $\eta > 0.7$.]{\includegraphics[width = 0.4\textwidth]{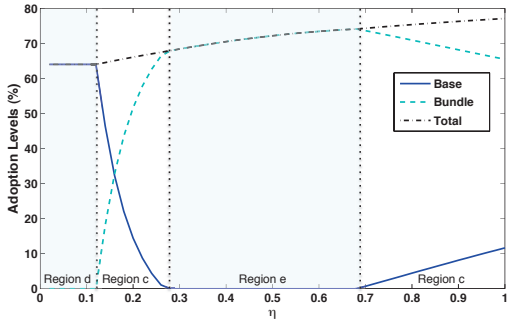} \label{fig:regionC}}%\hspace{-0.04\textwidth}
 \subfloat[Adoption as $\eta$ varies, $\overline{x}_1 = 0$ for $\eta > 0.1$.]{\includegraphics[width = 0.4\textwidth]{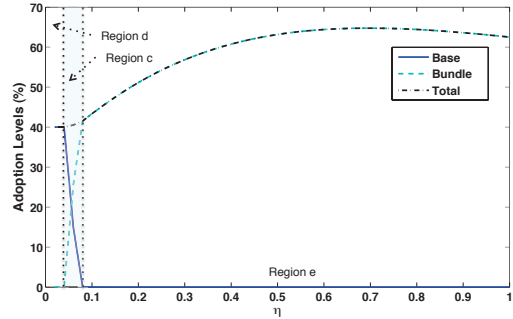} \label{fig:regionDEC}}%\hspace{-0.04\textwidth}
 \caption{As the supplemental technology's coverage area $\eta$ increases, (a) $\overline{x}_{1 + 2}$ decreases for large $\eta$, while (b) total adoption may also decrease.  Parameter values are (a) $q_1 = 200$, $q_2 = 250$, $\gamma_1 = 50$, $\gamma_2 = 20$, $p = 40$, $\Delta = 10$ and (b) $q_1 = 100$, $q_2 = 300$, $\gamma_1 = 50$, $\gamma_2 = 100$, $p = 40$, $\Delta = 10$.}
 \label{fig:behavior_sims}
 \vspace{-0.2in}
 \end{figure*}
% We consider possible adoption outcomes when the coverage factor and the prices offered vary.
 
 We first consider the adoption behavior for a range of coverage factors $(\eta)$, e.g., a {\color{red} WSP} that increases its WiFi coverage to offload more traffic from 3G, but cannot change its access prices due to the presence of a competitor.  Figure \ref{fig:regionC} shows the equilibrium adoption levels for a set of exogenous system parameters.  {\color{red} At {\color{blue} very small values of $\eta$ $\left(< 0.1\right)$}, the adoption levels lie in region d of Table \ref{tab:equilibria} and users adopt the bundle only if Technology 2 has sufficient coverage, i.e., $\eta \geq 0.1$.} At {\color{blue} large values of $\eta$  $\left(> 0.7\right)$}, adoption $\overline{x}_{1 + 2}$ of the bundled technologies decreases with $\eta$, even though the coverage area increases.  We can explain this phenomenon by noting that {\color{blue} as $\eta$ increases, two effects are present. First, each adopter of Technologies (1 + 2) can offload a larger amount of traffic to Technology 2. The total amount of traffic offloaded, $\eta \overline{x}_{1 + 2}$, therefore increases, leading to more congestion on Technology 2. Lower-end users then cease adopting the bundled technologies (i.e., the user valuation thresholds $\theta_{(1 + 2,1)}$ and $\theta_{(1 + 2,0)}$ for adopting the bundle will increase) and adopt only the base technology instead. Second, increasing $\eta$ allows more traffic to be offloaded from Technology 1 to Technology 2, decreasing the congestion on Technology 1 and inducing some non-adopters to begin adopting Technology 1: the valuation threshold $\theta_{(1,0)}$ for adoption of Technology 1 decreases. Thus, though adoption of the bundled technologies will decrease, adoption of Technology 1 will increase, and the total adoption $\overline{x}_1 + \overline{x}_{1 + 2}$ will also increase due to the movement of non-adopters towards Technology 1. % {\color{blue} We see this behavior for $\eta > 0.7$ in Fig. \ref{fig:regionC}.}
 
 Figure \ref{fig:regionDEC} shows an example in which the second effect is not present, i.e., non-adopters still prefer to adopt \emph{neither} technology. In this case, the increased congestion on Technology 2 induces some bundled adopters to adopt neither technology, rather than adopting Technology 1 $\left(\theta_{(1,0)} > \theta_{(1 + 2,0)} > \theta\right)$.
 % , i.e., the threshold for adopting the bundled Technologies (1 + 2) rather than no technology is larger than the threshold for adopting Technology 1.
Non-adopters still prefer to adopt neither technology, and in fact no users adopt the base technology $\left(\overline{x}_1 = 0\right)$.  Thus, $\overline{x}_1 + \overline{x}_{1 + 2} = \overline{x}_{1 + 2}$ and the total adoption may decrease:}
 
 \vspace{0.05in}
 \begin{result}\label{prop:x2eta}
\emph{Increasing the coverage factor $\eta$ can cause the total adoption $\overline{x}_1 + \overline{x}_{1 + 2}$ to decrease.}

Suppose that the coverage factor $\eta$ increases from $\eta_0$ to $\eta_1$. Then the total adoption at the new equilibrium adoption levels (for which $\eta = \eta_1$) is larger than the total adoption at the former equilibrium (for which $\eta = \eta_0$) if and only if, for any $\eta\in\left[\eta_0, \eta_1\right]$, the equilibrium adoption outcome is such that no users adopt the base technology $\left(\overline{x}_1 = 0\right)$, some users adopt the bundled technologies $\left(\overline{x}_{1 + 2} > 0\right)$, and
 \begin{align}
&(1 - \eta)^2\gamma_1q_1 + (1 - \eta^2)\gamma_1q_2 + \left(p + \Delta\right)\left(q_2 - q_1 - 2(1 - \eta)\gamma_1\right) \nonumber \\ &+ \eta\gamma_2\big((\eta - 2)q_1 - \eta q_2 + 2\left(p + \Delta\right)\big) < 0.
\label{eq:eWFeta}
\end{align}
%Suppose that no users adopt Technology 1 ($\overline{x}_1 = 0$), and that some, but not all, adopt both technologies ($\overline{x}_{1 + 2} \in (0,1)$).  Then the total adoption, i.e., $\overline{x}_{1 + 2}$, decreases with $\eta$ if
% \begin{align}
%&(1 - \eta)^2\gamma_1q_1 + (1 - \eta^2)\gamma_1q_2 + \eta(\eta - 2)\gamma_2q_1 - \eta^2\gamma_2q_2 \nonumber \\ &+ \left(p + \Delta\right)\left(q_2 - q_1 - 2(1 - \eta)\gamma_1 + 2\eta\gamma_2\right) < 0.
%\label{eq:eWFeta}
%\end{align}
%If some users adopt Technology 1, some adopt both technologies, and some neither ($\overline{x}_1 > 0$, $\overline{x}_{1 + 2} > 0$ and $\overline{x}_1 + \overline{x}_{1 + 2} < 1$), then total adoption $\overline{x}_1 + \overline{x}_{1 + 2}$ increases with $\eta$. % Adoption of Technologies (1 + 2) may either increase or decrease with $\eta$.
\end{result}
\vspace{0.07in}

We note that the left-hand side of (\ref{eq:eWFeta}) is decreasing in $\eta\gamma_2$ for sufficiently small prices: Technology 2's throughput degradation coefficient $\gamma_2$, multiplied by $\eta$, must be sufficiently high for $\overline{x}_{1 + 2}$ to decrease as $\eta$ increases.  Without {\color{red} such a} loss of utility from congestion on Technology 2, users will continue to adopt the bundled Technologies (1 + 2) for large $\eta$. This effect, however, may be outweighed by a large price $p + \Delta$ for the bundle, {\color{red} which increases} the positive $\left(p + \Delta\right)\left(q_2 - q_1\right)$ term. If the {\color{red} Technology 2 has a much higher intrinsic quality, i.e.,} $\left(q_2 - q_1\right)$ is large, users may adopt the bundled technologies as $\eta$ increases even if Technology 2 has a {\color{red} large $\gamma_2$}.
 
% For different parameter values, overall adoption $\overline{x}_1 + \overline{x}_{1 + 2}$ may decrease, as shown in Fig. \ref{fig:regionDEC}.  For these $q$ and $\gamma$ values, the utility of the 3G network is too low for any adoption to occur, so that the total adoption is limited to the (decreasing) WiFi adoption.
 
Another interesting feature of Fig. \ref{fig:regionDEC} is the {\color{blue} rapid} switch from all users adopting the base technology to all users adopting the bundled technologies when $\eta < 0.08$.  In this example, the access price $\Delta$ of Technology 2 is relatively low, as is its throughput degradation coefficient $\gamma_2$ when compared to $q_2$.  Thus, as $\eta$ increases slightly, the utility of adopting Technologies (1 + 2) increases quickly: the user need not pay much more for Technology 2, which provides higher quality service with relatively little throughput degradation.  Many users then adopt the supplemental technology in addition to the base one.  As $\eta$ grows further to 0.08, the utility of adopting {\color{blue} both technologies exceeds} that of adopting only the base technology $\big(\theta_{(1 + 2,1)} < \theta_{(1,0)}\big)$, save for those users who adopt neither technology due to low valuation levels $\theta$.
 
 \begin{figure}
 \centering
\includegraphics[width = 0.4\textwidth]{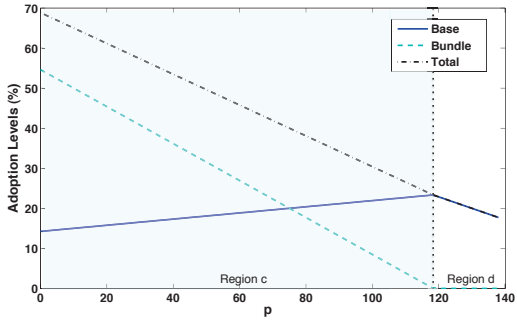}
 \vspace{-0.15in}
 \caption{As the base technology's access price $p$ increases, the base technology's adoption may increase (parameters: $q_1 = 200$, $q_2 = 225$, $\gamma_1 = 150$, $\gamma_2 = 50$, $\Delta = 30$, $\eta = 0.5$).}
 \label{fig:p3G}
 \vspace{-0.2in}
 \end{figure}

% Optimal prices
\begin{table*}[t]
\centering
\renewcommand{\arraystretch}{2.1}
\caption{Revenue-maximizing prices assuming equilibrium adoption levels in regions a-g (cf. Tables \ref{tab:regions}-\ref{tab:c}).}
\label{tab:revenue}
\vspace{-0.1in}
\begin{tabular}{|c|c|c|c|}
\hline
& $p^\ast$ & $\Delta^\ast$ & Revenue \\ \hline
a & $< -\left(1 - \eta\right)\gamma_1$ & $-(1 - \eta)^2 \gamma_1 - \eta^2\gamma_2 - p$ & $-(1 - \eta)^2 \gamma_1 - \eta^2\gamma_2$ \\ \hline
b$^\ast$ & $\frac{-\eta\gamma_1\gamma_2 + \left(1 - \frac{\eta}{2}\right)\left(q_1 - q_2\right)}{\eta\left(\gamma_1 + \gamma_2\right) - q_1 + q_2}$ & $\frac{\eta\left(q_2 - q_1\right)}{2}$ & $\frac{\frac{\eta}{4}\left(q_1 - q_2\right)^2 + (1 - \eta)\gamma_1\left(q_1 - q_2\right) - \eta\gamma_1\gamma_2}{\eta\left(\gamma_1 + \gamma_2\right) + q_2 - q_1}$ \\ \hline
c$^\dagger$ & $\frac{q_1}{2}$ & $\frac{\eta\left(q_2 - q_1\right)}{2}$ & $\frac{q_1^2\eta\gamma_2 + q_2^2\eta\gamma_1 + q_1^2\left(q_2 - q_1\right) + \eta q_1\left(q_1 - q_2\right)^2}{4\left(\gamma_1q_2 + \eta\gamma_1\gamma_2 + q_1\left(q_2 - q_1 + \eta\gamma_2 - (1 - \eta)\gamma_1\right)\right)}$ \\ \hline
d & $\frac{q_1}{2}$ & $\geq \eta\left(q_2 - q_1 + \frac{\gamma_1q_1}{2\left(q_1 + \gamma_1\right)}\right)$ & $\frac{q_1^2}{4(q_1 + \gamma_1)}$ \\ \hline
e & $\frac{(1 - \eta)q_1 + \eta q_2}{2} - \Delta$ & $\leq p\left(\frac{\eta q_2 + \eta^2\gamma_2}{q_1 + (1 - \eta)\gamma_1} - \eta\right) - \frac{\eta^2q_1\gamma_2 - \eta(1 - \eta)\gamma_1q_2}{q_1 + (1 - \eta)\gamma_1}$ & $\frac{\left((1 - \eta)q_1 + \eta q_2\right)^2}{4\left((1 - \eta)q_1 + \eta q_2 + (1 - \eta)^2\gamma_1 + \eta^2\gamma_2\right)}$ \\ \hline
f & $-\gamma_1$ & $\geq \eta\left(q_2 + \gamma_1 - q_1\right)$ & $-\gamma_1$ \\ \hline
g & $> q_1$ & $\geq \eta(q_2 - q_1)$ & 0 \\ \hline
\multicolumn{4}{|c|}{$^\ast$If $2(1 - \eta)\gamma_1 - 2\eta\gamma_2 > q_2 - q_1$, the revenue-maximizing prices for region b are instead: $p^\ast = \frac{(1 - \eta)\gamma_1\left(-\gamma_1 + q_1 - q_2\right)}{\eta\left(\gamma_1 + \gamma_2\right) + q_2 - q_1}$,} \\ 
\multicolumn{4}{|c|}{$\Delta^\ast = (1 - \eta)\gamma_1 - \eta\gamma_2$,  revenue = $\frac{-\eta\gamma_1\gamma_2 + \left((1 - \eta)^2\gamma_1 + \eta^2\gamma_2\right)\left(q_1 - q_2\right) - \eta\left((1 - \eta)\gamma_1 - \eta\gamma_2\right)^2}{\eta\left(\gamma_1 + \gamma_2\right) + q_2 - q_1}$} \\ \hline
\multicolumn{4}{|c|}{$^\dagger$If $\eta\gamma_2 q_1 \leq (1 - \eta)\gamma_1 q_2$, then at the optimal prices $\overline{x}_1 = 0$ and the equilibrium lies in region e.} \\ \hline
% \multicolumn{4}{|c|}{The optimal $\Delta$ for region e is $\leq p\left(\frac{\eta q_2 + \eta^2\gamma_2}{q_1 + (1 - \eta)\gamma_1} - \eta\right) - \frac{\eta^2q_1\gamma_2 - \eta(1 - \eta)\gamma_1q_2}{q_1 + (1 - \eta)\gamma_1}$} \\ \hline
\end{tabular}
\vspace{-0.1in}
\end{table*}
% From Appendix C
\begin{figure*}
\centering
\subfloat[$\beta$ distribution parameters $(\alpha,\beta) = (5,2)$.]{\includegraphics[width = 0.33\textwidth]{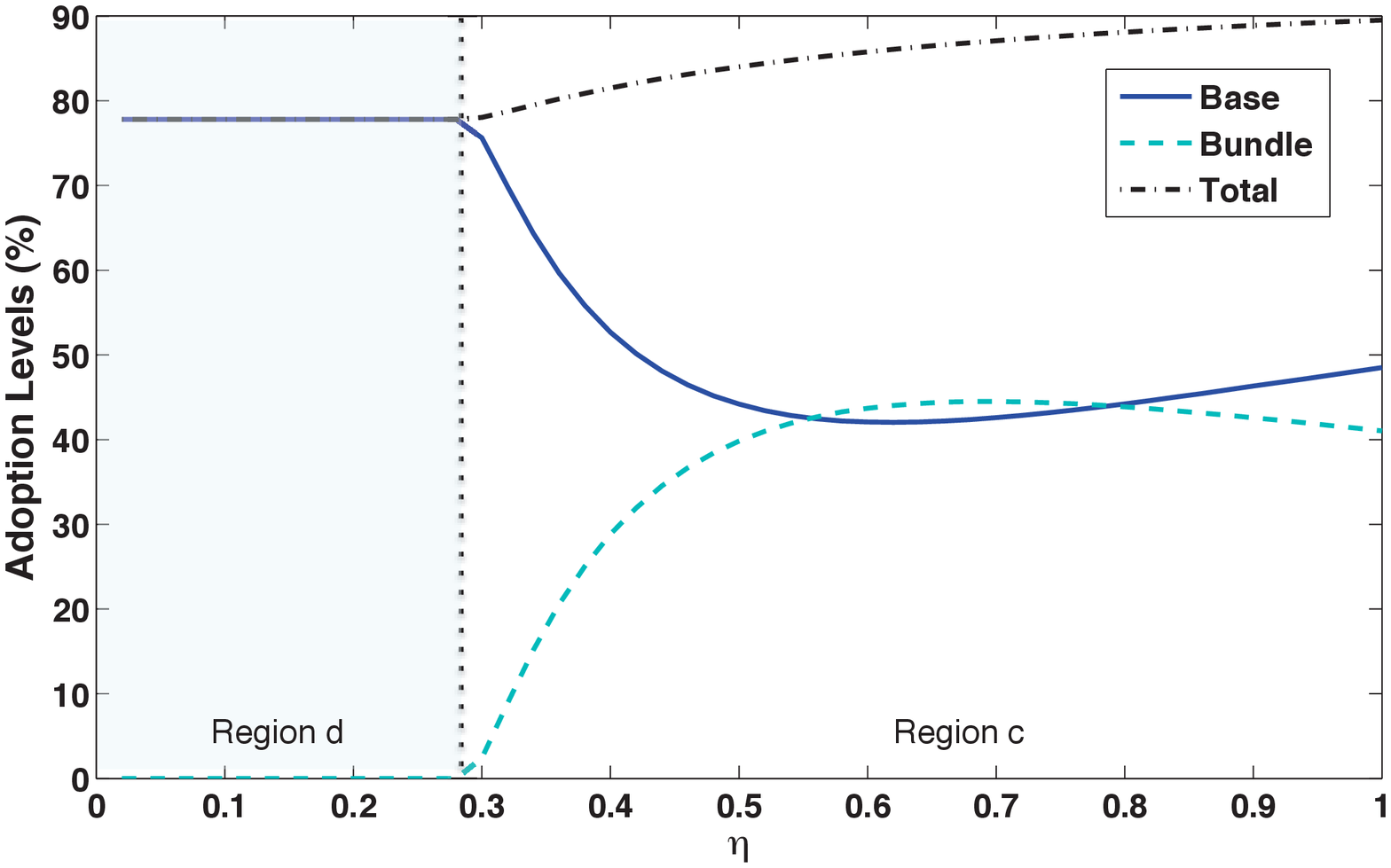}}\hspace{-0.015\textwidth}
\subfloat[$\beta$ distribution parameters $(\alpha,\beta) = (2,2)$.]{\includegraphics[width = 0.33\textwidth]{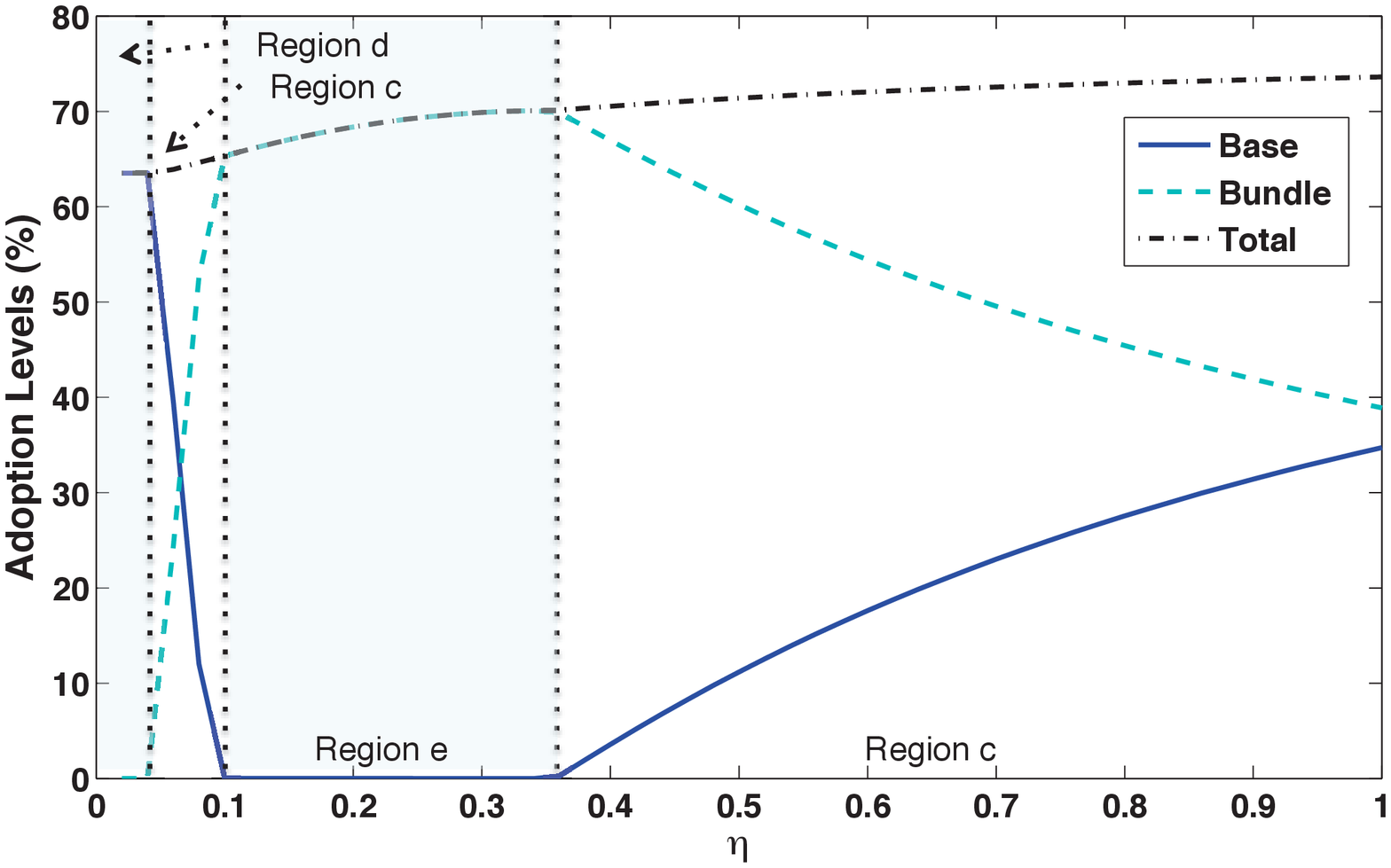}}\hspace{-0.015\textwidth}
\subfloat[$\beta$ distribution parameters $(\alpha,\beta) = (1,3)$.]{\includegraphics[width = 0.33\textwidth]{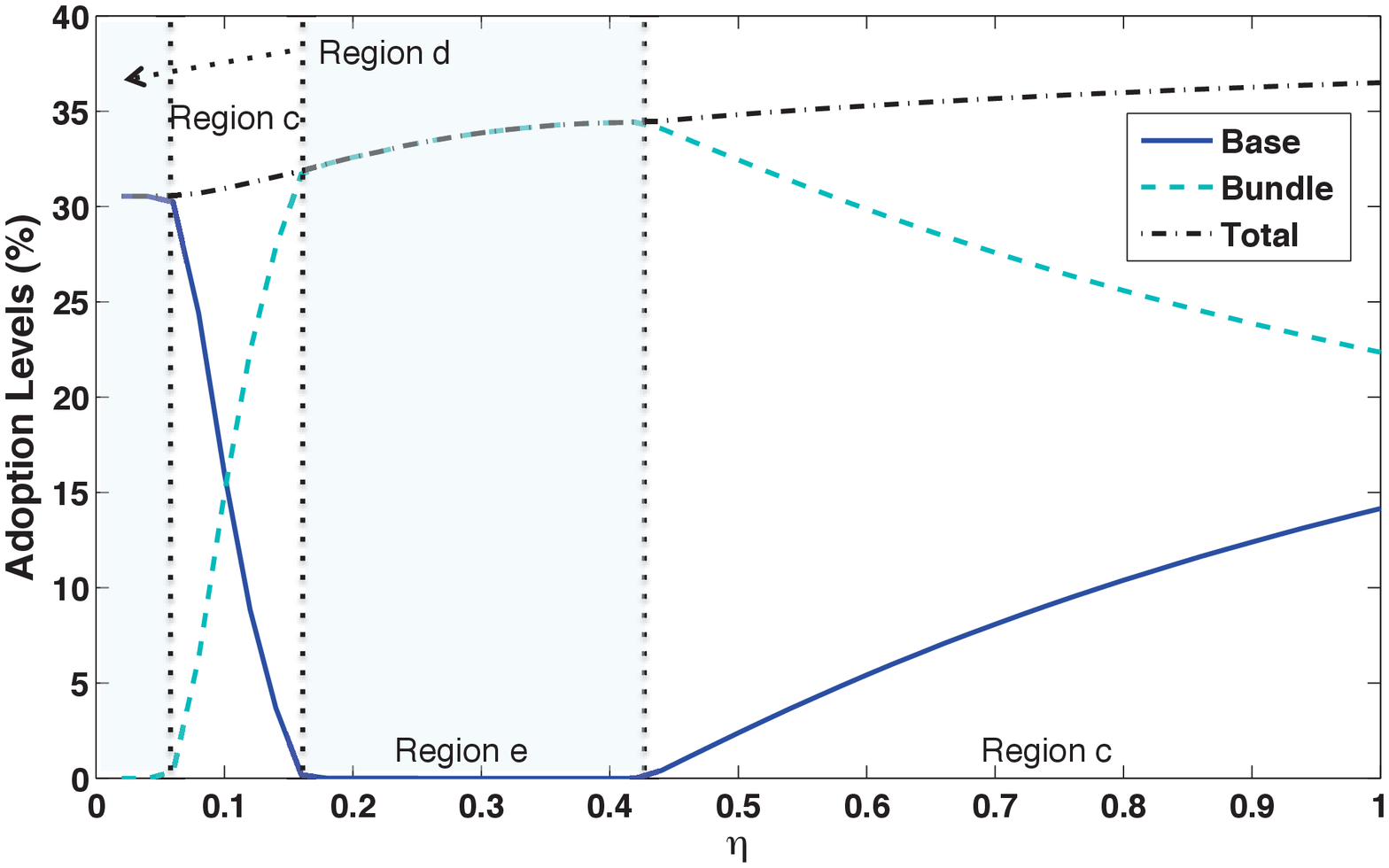}}
\caption{Equilibrium adoption levels as the coverage factor $\eta$ varies for different $\beta$ distributions of the user {\color{blue} valuations} $\theta$.  System parameters are (a) $q_1 = 150$, $q_2 = 250$, $\gamma_1 = 50$, $\gamma_2 = 150$, $p= 50$, $\Delta = 40$; (b) and (c) $q_1 = 200$, $q_2 = 280$, $\gamma_1 = 50$, $\gamma_2 = 150$, $p= 50$, $\Delta = 5$.}
\label{fig:behavior_nonuniform}
\vspace{-0.2in}
\end{figure*}

\subsection{Effect of the Base Access Price}\label{sec:prices}

We next consider adoption behaviors for a fixed Technology 2 access price $\Delta$ and coverage factor $\eta$.  For instance, the {\color{red} WSP} may increase the base technology's access price $p$ in an attempt to induce {\color{blue} some} users to leave the network.  However, we find that in some cases, this approach can backfire: increasing $p$ may instead increase Technology 1's adoption $\overline{x}_1$.  Figure \ref{fig:p3G} shows an example; we note that though $\overline{x}_1$ increases {\color{blue} for $p > 118$}, the total adoption $\overline{x}_1 + \overline{x}_{1 + 2}$ decreases.  We then see that:

\vspace{0.05in}
 \begin{result}\label{prop:p3G}
\emph{Increasing the base price $p$ need not decrease the base technology's adoption $\overline{x}_1$.}

{\color{red} Suppose that $p$ increases from $p_0$ to $p_1$.} Then the base technology's equilibrium adoption when $p = p_1$ is larger than that at which $p = p_0$ if and only if the equilibrium adoption outcome for any $p\in \left[p_0, p_1\right]$ is such that some users adopt Technology 1 and some adopt {\color{blue} Technologies 1 + 2 (i.e., $\overline{x}_1 > 0$ and $\overline{x}_{1 + 2} > 0$ as in region c)}, and
\begin{equation}
q_2 - q_1 < (1 - \eta)\gamma_1 - \eta\gamma_2.
\label{eq:p3G}
\end{equation}
\end{result}
\vspace{0.07in}
 
The quantity $(1 - \eta)\gamma_1$ on the right side of (\ref{eq:p3G}) represents the throughput improvement on Technology 1 due to users of Technologies (1 + 2) offloading traffic onto Technology 2, while $-\eta\gamma_2$ represents the throughput degradation on Technology 2 due to congestion. {\color{red} Thus, (\ref{eq:p3G})} indicates that Technology 1's adoption $\overline{x}_1$ increases with $p$ if the quality differential $(q_2 - q_1)$ from adoption of Technology 2 is outweighed by the marginal savings in throughput degradation from adopting only Technology 1 $\left((1 - \eta)\gamma_1 - \eta\gamma_2\right)$.  Users will then choose to adopt Technology 1, rather than the bundled technologies: the valuation threshold $\theta_{(1 + 2,1)}$ for adopting the bundled technologies will increase. This increase may be faster than that of {\color{red} $\theta_{(1,0)}$, i.e.,} more users will drop from the bundled technology to Technology 1 than the users who cease adopting Technology 1, leading to a net increase in $\overline{x}_1$. As shown in Fig. \ref{fig:p3G}, once the number of users of Technologies (1 + 2) has gone to zero, {\color{blue} the dynamics transition from region c to region d and users no longer switch from adopting Technologies (1 + 2) to adopting Technology 1. The base technology's adoption $\overline{x}_1$ then begins to decrease with Technology 1's price $p$.} {\color{red} The threshold price at which this transition occurs is that at which $\overline{x}_{1 + 2} = 0$ {\color{blue} in region c, i.e.,} $p(\Delta,\eta) = q_2 + q_1\left(q_2 - q_1\right)/\gamma_1 - \Delta\left(\gamma_1 + q_1\right)/(\eta\gamma_1)$.}

{\color{red} \subsection{Nonuniform Valuation Distributions}\label{sec:nonuniform_theta}

In {\color{red} Figs. \ref{fig:behavior_nonuniform} and \ref{fig:behavior_x10}}, we show adoption behaviors for fixed system parameters when the user valuations $\theta$ are not uniformly distributed.  We consider three different distributions of $\theta${\color{blue}, chosen as different $\beta$ distributions,} and investigate the equilibrium adoption levels as the coverage factor $\eta$ varies.

We first note that despite the nonuniform heterogeneity, equilibria exist for each value of $\eta$ simulated. In fact, as in Fig. \ref{fig:regionC} with a uniformly distributed variable $\theta$, {\color{blue} in Fig. \ref{fig:behavior_nonuniform}a} we observe that as the coverage increases, adoption $\overline{x}_{1 + 2}$ of the bundled technologies decreases, while total adoption $\overline{x}_1 + \overline{x}_{1 + 2}$ increases.  Our simulation in Fig. \ref{fig:behavior_nonuniform}b even replicates Fig. \ref{fig:regionDEC}'s {\color{blue} rapid} switch from predominant adoption of only Technology 1 ($\eta < 0.04$) to predominant adoption of only Technologies (1 + 2). In Fig. \ref{fig:behavior_x10}, we present an example in which $\overline{x}_1 = 0$ for $\eta > 0.12$; then as the adoption of Technologies (1 + 2) decreases, so does the total adoption. This adoption behavior is qualitatively comparable to that observed in Fig. \ref{fig:regionDEC}.

\section{Revenue and Profit Maximization}\label{sec:profit}

{\color{blue} Having used our model to identify and characterize a range of equilibrium adoption behaviors, we now use the insights} derived in Sections \ref{sec:usage} and \ref{sec:guidelines} to analyze the implications for {\color{red} WSP} profit and revenue. We first consider the {\color{red} WSP}'s revenue in Section \ref{sec:revenue}, deriving the revenue-maximizing prices and investigating the corresponding adoption behavior.  We then introduce two additional factors in Section \ref{sec:opt_profit}, {\color{red} WSP} savings from offloading traffic and the cost of deploying the supplemental technology. We use empirical usage data to estimate these costs and the resulting adoption behaviors.

\subsection{Revenue Maximization}\label{sec:revenue}

We {\color{blue} first consider} the behavior of Technologies (1 + 2)'s equilibrium adoption $\overline{x}_{1 + 2}$ {\color{red} as $\eta$ varies} and the {\color{red} WSP} chooses prices so as to maximize its revenue $p\left(\overline{x}_1 + \overline{x}_{1 + 2}\right) + \Delta\overline{x}_{1 + 2}$.  We use Tables \ref{tab:equilibria} and \ref{tab:c}'s expressions for the equilibrium $\overline{x}_1$ and $\overline{x}_{1 + 2}$ to find the revenue-maximizing prices $p^\ast$ and $\Delta^\ast$ at each possible equilibrium (Table \ref{tab:revenue}).  To emphasize their dependence on price, in the rest of this section the notation $\overline{x}_1^\ast = \overline{x}_1\big(p^\ast, \Delta^\ast\big)$ and $\overline{x}_{1 + 2}^\ast = \overline{x}_{1 + 2}\big(p^\ast, \Delta^\ast\big)$ denotes the equilibrium adoption levels at the optimal prices $p^\ast$ and $\Delta^\ast$.
 
We see from Table \ref{tab:revenue} that the {\color{red} WSP} earns non-positive revenue if it maximizes its revenue at equilibria in regions a, f, or g. In region g, there is no adoption, leading to zero revenue. In regions a and f, the {\color{red} WSP} observes full adoption, and as observed in Section \ref{sec:guidelines}, it must therefore offer \emph{negative} prices $p$ and $\Delta$, leading to negative revenue. In region b, which also has full adoption, the negative price of Technology 1 ($p^\ast$) may be offset by a sufficiently positive price $p^\ast + \Delta^\ast$ for Technologies (1 + 2), so that the overall revenue is positive. This (possibly) positive revenue, however, is exceeded by that in other regions. {\color{blue} We therefore focus on regions b--e.}
Using Table \ref{tab:revenue}, we can prove that revenue is greatest under partial adoption of both technologies:

\vspace{0.05in}
\begin{proposition}\label{prop:rev_partial}
{\color{red} WSP} revenue is maximized with partial, but not full, adoption ($\overline{x}_1^\ast + \overline{x}_{1 + 2} ^\ast < 1$) and full coverage ($\eta = 1$).
\end{proposition}
\vspace{0.07in}

If the {\color{red} WSP} is not free to vary its coverage factor, then we can also derive conditions under which no users adopt the base technology at the revenue-maximizing prices:

\vspace{0.05in}
\begin{result}\label{prop:revenuemax}
For any fixed $\eta$, some users adopt the base technology at the revenue-maximizing prices if and only if
\begin{equation}
\eta\frac{\gamma_2}{\gamma_1} > (1 - \eta)\frac{q_2}{q_1}. \label{eq:cond}
\end{equation}
Thus, if (\ref{eq:cond}) is not satisfied, the {\color{red} WSP} can eliminate data plans {\color{blue} for only the base technology, and only offer bundled plans}.
%the revenue-maximizing equilibrium adoption levels lie in region c of Table \ref{tab:regions}: some users adopt Technology 1, some adopt Technologies (1 + 2), and some adopt neither technology $\big(\overline{x}_1\big(p^\ast, \Delta^\ast\big) > 0$, $\overline{x}_{1 + 2}\big(p^\ast, \Delta^\ast\big) > 0$, $\overline{x}_1\big(p^\ast, \Delta^\ast\big) + \overline{x}_{1 + 2}\big(p^\ast, \Delta^\ast\big) < 1\big)$.  If (\ref{eq:cond}) does not hold, then no users adopt Technology 1, but some adopt Technologies (1 + 2) $\left(\overline{x}_1\big(p^\ast, \Delta^\ast\big) = 0.\;\overline{x}_{1 + 2}\big(p^\ast, \Delta^\ast\big) > 0\right)$.
\end{result}
\vspace{0.07in}

The condition in {\color{red} Result} \ref{prop:revenuemax} can be interpreted as stating that when the quality $q_1$ of Technology 1 is sufficiently high and the marginal throughput degradation $\gamma_1$ sufficiently low relative to Technology 2, then for {\color{blue} a large coverage factor $\eta$}, some users will adopt Technology 1 at the optimal prices. % A large $q_1$ and small $\gamma_1$ imply that users value Technology 1 and are less affected by its congestion level, while a {\color{red} large $\eta$} indicates that a significant amount of traffic can be offloaded {\color{blue} from Technology 1} to Technology 2, relieving congestion on Technology 1.

Under these conditions, the adoption $\overline{x}_{1 + 2}\big(p^\ast, \Delta^\ast\big)$ of Technologies (1 + 2) will decrease {\color{blue} as $\eta$ increases}, as shown in Fig. \ref{fig:maxrev_eta}'s example.\footnote{We note that the overall adoption levels in Fig. \ref{fig:maxrev_eta} are low when compared with those of Fig. \ref{fig:behavior_sims}.  With different parameters (e.g. $q_i$ and $\gamma_i$ values), the overall adoption levels may change; we use the ones here to reflect current smartphone penetration rates in the U.S. \cite{gigaom}.}  As in Fig. \ref{fig:regionC}, {\color{blue} two effects are present. First, the total volume of traffic $\eta \overline{x}_{1 + 2}^\ast$ offloaded onto Technology 2 increases with $\eta$, increasing Technology 2's adoption threshold $\theta_{(1 + 2,1)}$ and driving more users to adopt only Technology 1 instead of the bundle. Second, overall adoption $\overline{x}_1\big(p^\ast, \Delta^\ast\big) + \overline{x}_{1 + 2}\big(p^\ast, \Delta^\ast\big)$ increases: the increase in $\eta$ leads to more traffic being offloaded from Technology 1 to Technology 2, reducing congestion on Technology 1 and inducing low-end non-adopters to adopt Technology 1. If this second effect is not present, then the overall adoption equals that of Technologies 1 + 2 and does not increase:}

\vspace{0.05in}
\begin{result}\label{prop:optx}
\emph{Increasing the coverage factor $\eta$ at the {\color{red} WSP}'s revenue-maximizing prices can cause the total adoption $\overline{x}_1\big(p^\ast, \Delta^\ast\big) + \overline{x}_{1 + 2}\big(p^\ast, \Delta^\ast\big)$ to decrease.}

Suppose that the coverage factor $\eta$ increases from $\eta_0$ to $\eta_1$. Then the total equilibrium adoption with revenue-maximizing prices when $\eta = \eta_1$ will {\color{red} only be smaller than that when $\eta = \eta_0$ if}, for any $\eta\in\left[\eta_0, \eta_1\right]$, the equilibrium adoption at the revenue maximizing prices is such that no users adopt the base technology $\left(\overline{x}_1^\ast = 0\right)$, or equivalently, (\ref{eq:cond}) does not hold.
\end{result}
\vspace{0.05in}
% From Section IV-C (nonuniform theta distribution)
\begin{figure}[t]
\vspace{-0.1in}
\centering
\includegraphics[width = 0.45\textwidth]{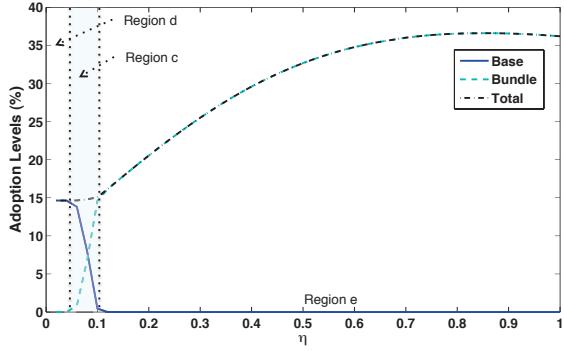}
\vspace{-0.15in}
\caption{Equilibrium adoption levels as the coverage factor $\eta$ increases, $\overline{x}_1 = 0$ for large $\eta$. User heterogeneity $\theta$ follows a $\beta$ distribution with parameters $(\alpha,\beta) = (1,3)$.  System parameters are $q_1 = 100$, $q_2 = 300$, $\gamma_1 = 50$, $\gamma_2 = 100$, $p = 40$, $\Delta = 10$.}
\vspace{-0.15in}
\label{fig:behavior_x10}
\end{figure}
% From Section IV-C (nonuniform theta distribution)
%\begin{figure}[t]
%\centering
%\includegraphics[width = 0.45\textwidth]{Figures/BetaDist.eps}
%\vspace{-0.15in}
%\caption{Different $\beta$ distributions used for user heterogeneity $(\theta)$ in Fig. \ref{fig:behavior_nonuniform}.}
%\vspace{-0.15in}
%\label{fig:betadist}
%\end{figure}
}

\begin{table*}
\centering
\renewcommand{\arraystretch}{1.7}
\caption{{\color{red} Adoption levels at the revenue-maximizing prices for fixed $\eta$.}}
\label{tab:adoption_revmax}
\vspace{-0.1in}
\begin{tabular}{|c|c|c|}
\hline
{\color{red} Condition} & {\color{red} $\overline{x}_1^\ast$} & {\color{red} $\overline{x}_{1 + 2}^\ast$} \\ \hline
{\color{red} $\eta\frac{\gamma_2}{\gamma_1} \leq (1 - \eta)\frac{q_2}{q_1}$} & {\color{red} $0$} & {\color{red} $\frac{(1 - \eta)q_1 + \eta q_2}{2\left((1 - \eta)q_1 + \eta q_2 + (1 - \eta)^2\gamma_1 + \eta^2\gamma_2\right)}$} \\ \hline
{\color{red} $\eta\frac{\gamma_2}{\gamma_1} > (1 - \eta)\frac{q_2}{q_1}$} & {\color{red} $\frac{(1 - \eta)\gamma_1 q_2 - \eta q_1\gamma_2}{2\left(-\gamma_1q_2 - \eta\gamma_1 \gamma_2 + q_1\left((1 - \eta)\gamma_1 - \eta\gamma_2 + q_1 - q_2\right)\right)}$} & {\color{red} $\frac{-\gamma_1 q_2 + q_1\left(q_1 - q_2\right)}{2\left(-\gamma_1q_2 - \eta\gamma_1 \gamma_2 + q_1\left((1 - \eta)\gamma_1 - \eta\gamma_2 + q_1 - q_2\right)\right)}$} \\ \hline
\end{tabular}
\vspace{-0.1in}
\end{table*}

% Example figure
\begin{figure}[t]
%\vspace{-0.05in}
\centering
\includegraphics[width = 0.45\textwidth]{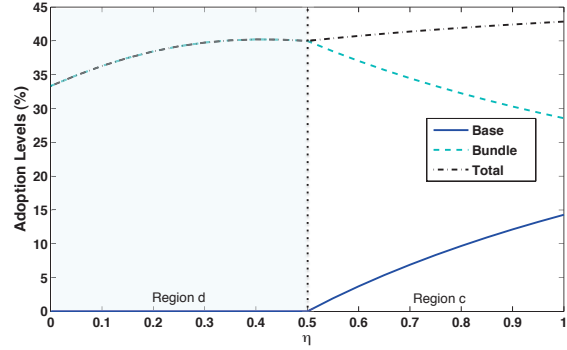}
\vspace{-0.15in}
\caption{Adoption levels for $\eta\in[0,1]$ and revenue-maximizing prices ($q_1 = 50$, $q_2 = 100$, $\gamma_1 = 50$, $\gamma_2 = 100$).  As $\eta$ increases, total adoption $\overline{x}_1\big(p^\ast, \Delta^\ast\big) + \overline{x}_{1 + 2}\big(p^\ast, \Delta^\ast\big)$ increases, driven by the increase in $\overline{x}_1$.}
\label{fig:maxrev_eta}
\vspace{-0.2in}
\end{figure}

% From the next section on adoption behaviors 
\begin{figure*}
\centering
\includegraphics[width = 0.95\textwidth]{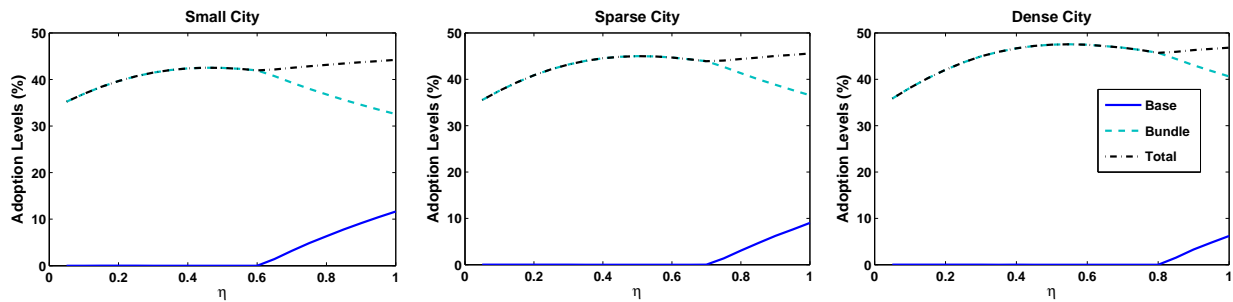}
\vspace{-0.15in}
\caption{Adoption levels for the profit-maximizing prices and fixed coverage factor $\eta$ ($q_1 = 50$, $q_2 = 100$, $\gamma_1 = 25$, $\gamma_2 = 50$) in different scenarios.  Nominal cost parameters are $\left(c_{\rm WF}, c_{\rm AP}\right) = (5.4, 6.2)$, $(10.6, 4.9)$ and $(15.9, 11.5)$ for the small, sparse, and dense cities respectively.  As $\eta$ increases past a threshold value, adoption $\overline{x}_{1 + 2}\big(p^\ast, \Delta^\ast\big)$ of Technologies (1 + 2) decreases despite the potential to offload more traffic {\color{blue} with a larger $\overline{x}_{1 + 2}\big(p^\ast, \Delta^\ast\big)$}.}
 \label{fig:etaall_opt}
 \vspace{-0.2in}
 \end{figure*}
 
We can further derive the exact adoption levels $\overline{x}_1^\ast$ and $\overline{x}_{1 + 2}^\ast$ as a function of $\eta$ from the proof of Result \ref{prop:revenuemax}. If (\ref{eq:cond}) holds, then the revenue-maximizing adoption levels occur in region c; if {\color{blue} not}, then they occur in region e. {\color{blue} From Tables \ref{tab:equilibria} and \ref{tab:revenue}, we then find the equilibrium adoption levels as in Table \ref{tab:adoption_revmax}.}
We also note that {\color{red} WSP} revenue increases with $\eta$. In fact, we observe from Table \ref{tab:revenue} (region c) that $\Delta^\ast = \eta\left(q_2 - q_1\right)/2$ increases with $\eta$; this increase in price offsets any decrease in adoption for Technologies (1 + 2), increasing the {\color{red} WSP} revenue.

\subsection{Optimizing Profit}\label{sec:opt_profit}
 
In addition to its revenue, {\color{red} WSP} profit includes its savings from offloading, less the cost of deploying a supplemental technology.\footnote{We assume that the deployment cost of the base technology is independent of the adoption levels, e.g., an already-deployed 3G network.}  Since these parameters depend on the market conditions, we consider three scenarios: a small city; {\color{blue} a large, sparsely populated city; and a large, more densely populated city}. We refer to the latter two cities as the ``sparse'' and ``dense'' cities. We {\color{blue} take the base technology to be 3G and the supplemental one to be WiFi, and collect empirical data to estimate their costs and savings from offloading. We then derive a mathematical model for the offloading savings and deployment costs and show how these costs affect the optimal coverage factor and adoption behaviors. We find that the adoption behavior with fixed coverage factor $\eta$ is qualitatively similar to the behavior when only revenue is maximized.} % Throughout the section, we take the , and we use cost parameters appropriate for these realizations.

\subsubsection{Trial Data}\label{sec:trial}

To estimate 3G and WiFi cost parameters, we gather 3G and WiFi usage data from 20 Android smartphones over six days. Since {\color{red} WSP} cost is driven by peak-hour traffic, we focus on usage when the 3G network is most heavily utilized \cite{BellLabs}.  Our goal is twofold: first, to estimate the fraction of 3G traffic that occurs at this peak time; and second, to estimate the probability of WiFi access at this time, given the overall WiFi access probability.  We can then estimate the amount of traffic that will be offloaded to WiFi at the peak time, given the WiFi adoption {\color{blue} level} and coverage factor.

We implemented a simple data monitoring app and released it to users in the United States.  In each hour, we recorded the volume of 3G and WiFi usage and WiFi base station IDs. % To ensure that our data represented a typical usage sample, users were given no incentives to adjust their data usage, aside from viewing the app itself.
We find that the probability of WiFi access in the hour of highest 3G usage is 82\% of the overall probability of WiFi access.  On average, 55\% of 3G traffic occurs in these peak hours, corroborating existing findings that 3G data usage exhibits severe peaks during the day \cite{BellLabs}. Moreover, the overall fraction of traffic on WiFi is fairly consistent across all participants, with an average of 71\% and variance of 9\%. This result helps validate our model's assumption of a single coverage factor affecting heterogeneous users' utilities from using the base and supplemental technologies. % Since we take the coverage factor $\eta$ as an optimization variable in the next section, we do not estimate its value.
 
\subsubsection{Cost Model}
 
We model the cost savings introduced by user offloading as a linear function of the amount offloaded during the peak hour, i.e., the marginal cost of peak traffic, multiplied by the amount offloaded \cite{BellLabs}.  The amount offloaded may be expressed as $C_{\rm WF}\eta\overline{x}_{1 + 2}$, where $C_{\rm WF}\eta$ is the probability that a {\color{blue} bundle} user has access to WiFi in the hour of peak 3G usage, multiplied by the {\color{blue} per-user} peak 3G usage. Multiplying by $\overline{x}_{1 + 2}$ then yields the {\color{blue} amount of traffic offloaded, normalized by the user population. The total savings from offloading is then $C_{\rm WF}\eta\overline{x}_{1 + 2}$ multiplied by the marginal savings}, which we denote as $c_{\rm WF}\eta \overline{x}_{1 + 2}$. As described in Appendix \ref{sec:costs}, we estimate $c_{\rm WF}$ using {\color{blue} our trial} data {\color{blue} to derive} $c_{\rm WF} = 5.4$, 10.6, or 15.8 for the small, sparse, and dense cities respectively.

We next consider the cost of deploying the supplemental technology.  We assume that the {\color{red} WSP}'s access point (AP) deployment in each type of city is such that the throughput degradation is the same function of the {\color{blue} traffic} on Technology 2's network (i.e., equal $\gamma_2$ values).  In more densely populated cities, the {\color{red} WSP} may utilize a denser AP deployment in order to accommodate the larger number of users in the sparse and dense cities.  These additional APs do not increase the geographical coverage area, but rather accommodate more users within the same area.

We model the cost of deployment as $c_{\rm AP}\eta$, a linear function of $\eta$; {\color{red} we use $c_{\rm AP} = 6.2$, 4.9, and 11.5 respectively for the small, sparse, and dense cities, as calculated in Appendix \ref{sec:costs}.} We thus maximize the {\color{red} WSP}'s total profit at equilibrium
\begin{equation}
p\left(\overline{x}_1 + \overline{x}_{1 + 2}\right) + \Delta\overline{x}_{1 + 2} + c_{\rm WF}\eta \overline{x}_{1 + 2} - c_{\rm AP}\eta
\label{eq:profit}
\end{equation}
with respect to the optimization variables $\eta$, $p$, and $\Delta$.  In the remainder of this section, we use $\eta^\ast$, $p^\ast$, and $\Delta^\ast$ to denote the optimal values of $\eta$, $p$, and $\Delta$, respectively; the corresponding adoption levels are denoted by $\overline{x}_1^\ast$ and $\overline{x}_{1 + 2}^\ast$.

% To calculate $c_{\rm AP}$ for use in simulation, we assume a standard AP range of 130 meters {\color{red} [citation needed]} and population densities of 2000, 5000, and 12000 people per square mile in the small city, large west coast city, and large east coast city respectively.  We suppose that the investment cost is \$1400, spread over one year \cite{wireless2020}.  Monthly operational costs are assumed to be small compared to the capital investment.  We use a standard AP range of 16 meters and find that $c_{\rm AP} = 6.2, 4.9, 11.5$ for the small city, large west coast, and large east coast city respectively.

\begin{figure}
\centering
\subfloat[Hourly 3G and WiFi usage.]{\includegraphics[width = 0.2\textwidth]{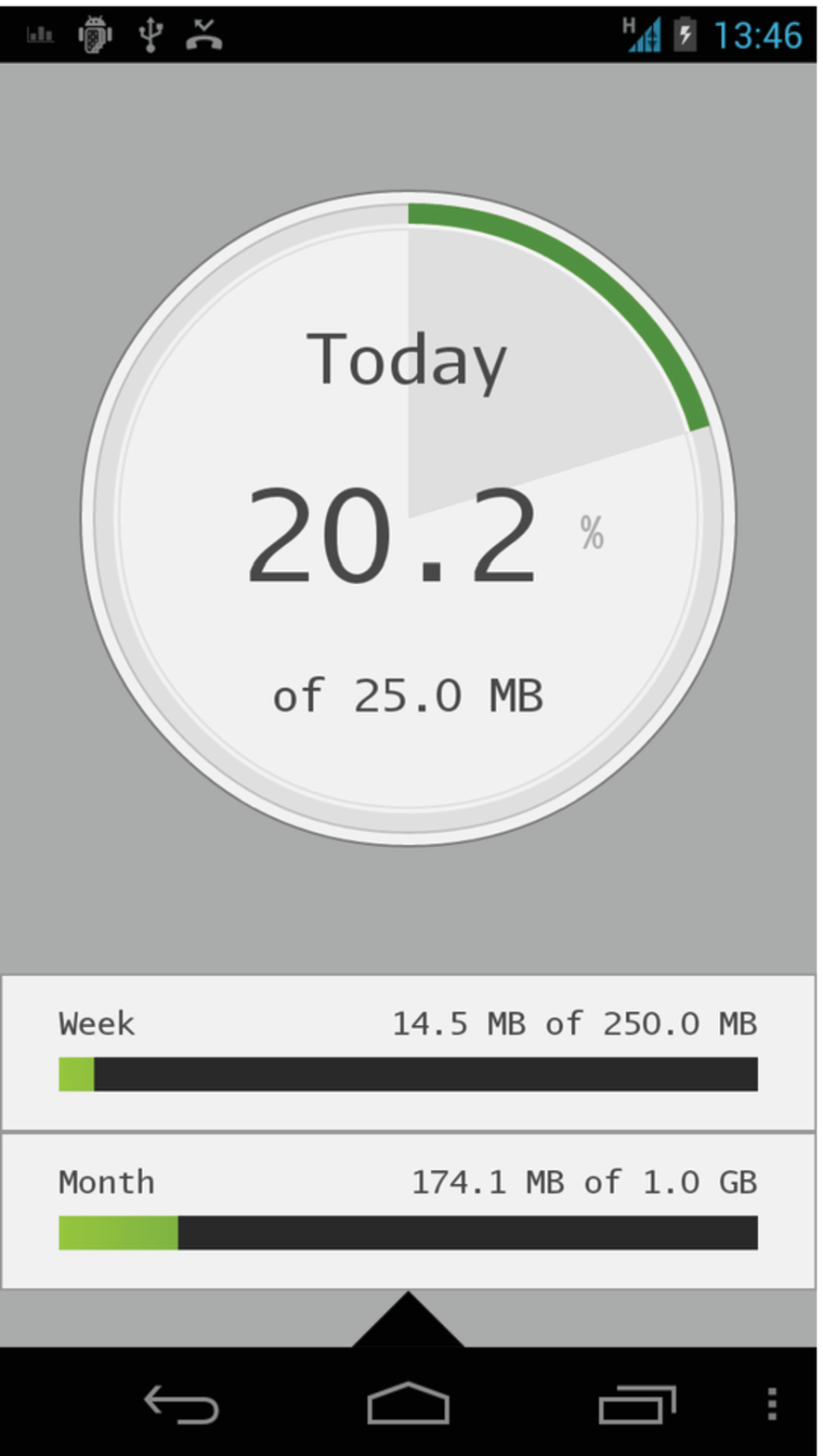}}\hspace{0.03\textwidth}
\subfloat[Location-specific usage.]{\includegraphics[width = 0.2\textwidth]{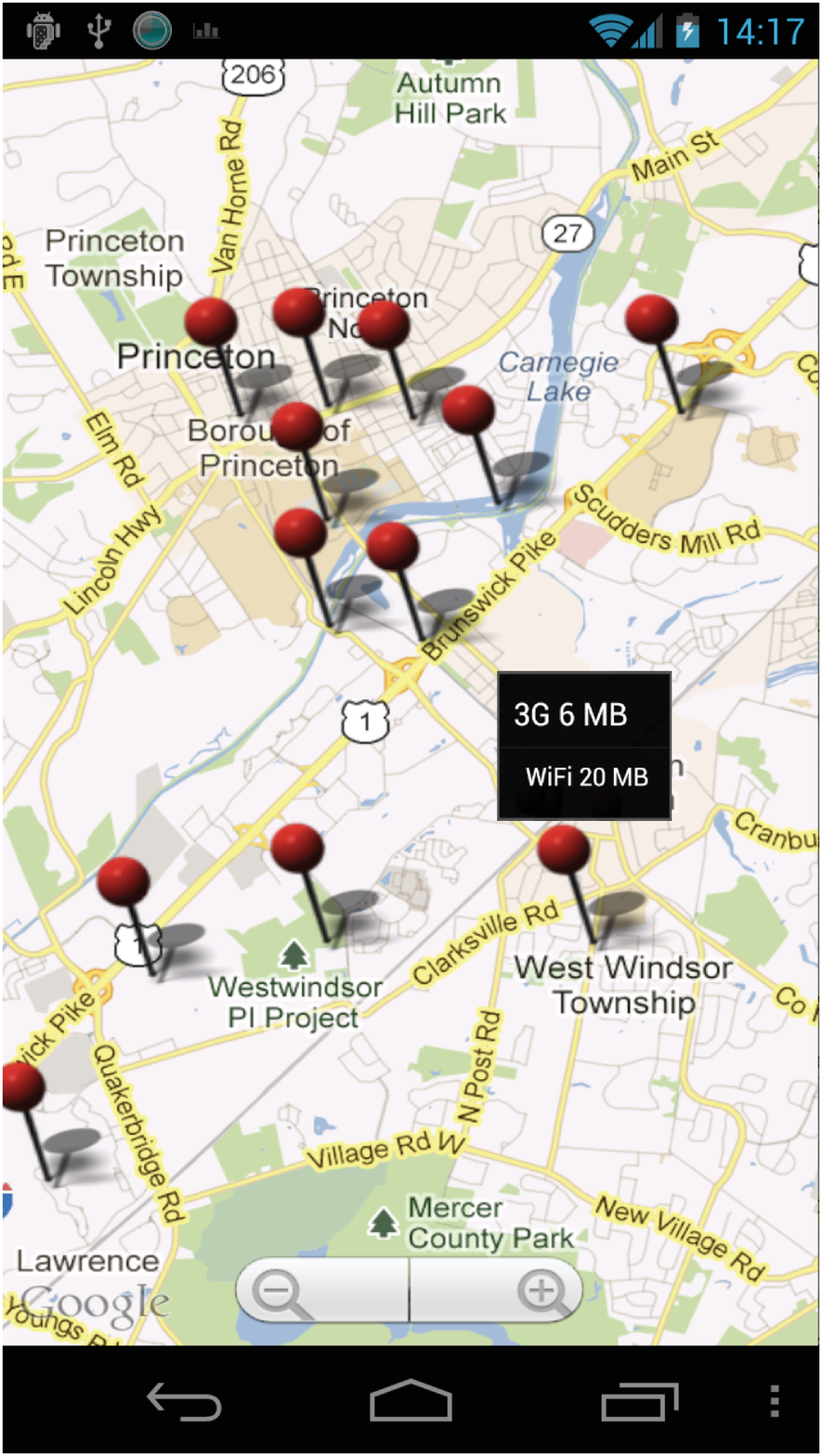}}
\caption{Screenshots of our usage monitoring app on the Android platform.}
\label{fig:screenshots}
\vspace{-0.15in}
\end{figure}
 
 \begin{figure}
\centering
\includegraphics[width = 0.4\textwidth]{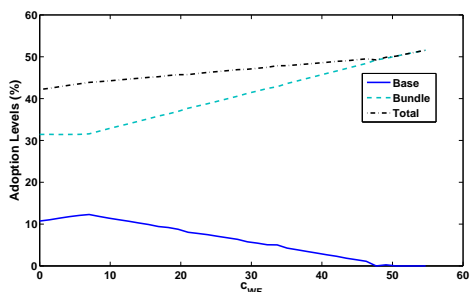}
\vspace{-0.15in}
\caption{Adoption levels for the profit-maximizing prices and coverage factor for the sparse city ($q_1 = 50$, $q_2 = 100$, $\gamma_1 = 25$, $\gamma_2 = 50$, $c_{\rm AP} = 4.9$).}
 \label{fig:costoffload_sparse}
 \vspace{-0.2in}
 \end{figure}
 
 \begin{figure}
 \centering
 \includegraphics[width = 0.4\textwidth]{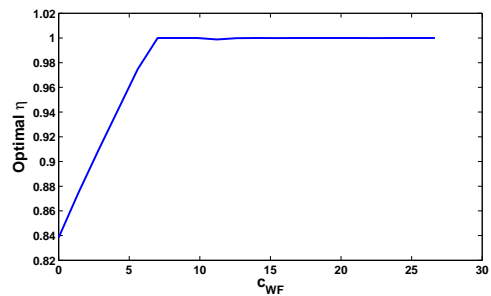}
 \vspace{-0.1in}
 \caption{Optimal coverage factors $\eta^\ast$ in Fig. \ref{fig:costoffload_sparse}.}
 \label{fig:coverage_offload}
 \vspace{-0.25in}
 \end{figure}

\begin{figure*}[ht]
\centering
\includegraphics[width = 0.95\textwidth]{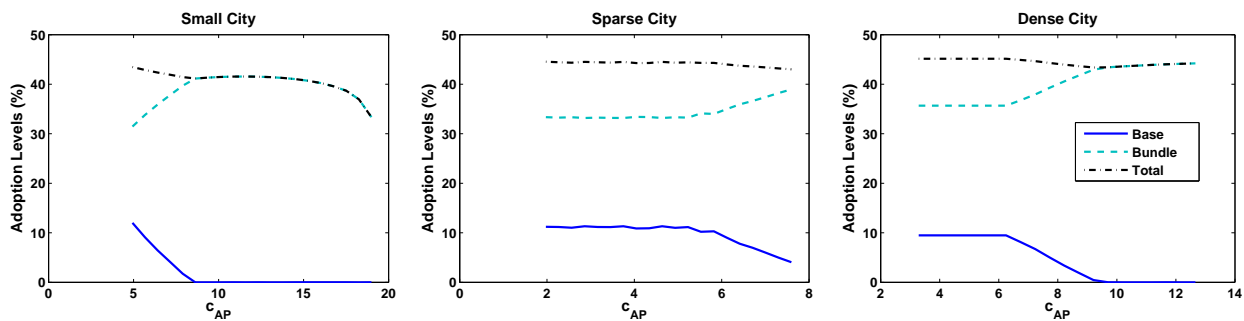}
\vspace{-0.15in}
\caption{Adoption levels for the profit-maximizing prices and coverage factor ($q_1 = 50$, $q_2 = 100$, $\gamma_1 = 25$, $\gamma_2 = 50$) in different scenarios.  Nominal cost parameters are $c_{\rm WF} = 5.4$, $10.6$, and $15.9$ for the small, sparse, and dense cities respectively.}
 \label{fig:costdeploy}
 \vspace{-0.2in}
 \end{figure*}
 
\subsubsection{Adoption Behaviors}
 
We first examine adoption behavior at the profit-maximizing prices $p^\ast$ and $\Delta^\ast$ for fixed coverage factor $\eta$ and cost parameters $c_{\rm AP}$ and $c_{\rm WF}$.  Since $\eta$ is fixed in this scenario, we see from (\ref{eq:profit}) that the only {\color{blue} cost} term in the optimization is $c_{\rm WF}\eta\overline{x}_{1 + 2}$; we would thus expect to observe higher adoption of the bundled technologies when profit is maximized, rather than revenue.

We show the profit-maximizing adoption levels in  Fig. \ref{fig:etaall_opt}; visually, their behavior is qualitatively similar to the revenue-maximizing adoption levels in Fig. \ref{fig:maxrev_eta}. In particular, even though the {\color{red} WSP} receives an additional offloading benefit from more users adopting the bundled technologies, adoption $\overline{x}_{1 + 2}^\ast$ of the bundled technologies still decreases for large $\eta$. We note, however, that as the marginal benefit from offloading $c_{\rm WF}$ increases ($c_{\rm WF}$ is smallest for the small city and largest for the dense city), the threshold value of $\eta$ at which $\overline{x}_{1 + 2}^\ast$ decreases grows. Thus, if $c_{\rm WF}$ is sufficiently large, the {\color{red} WSP}'s prices will be such that adoption of the bundled technologies never decreases as the coverage factor $\eta$ increases.
% we see that as $\eta$ increases, adoption of Technologies (1 + 2) first increases and then decreases.  For smaller values of $\eta$, the {\color{red} WSP} induces users to adopt Technology 2 and offload traffic onto this network.  As $\eta$ grows, however, the {\color{red} WSP} allows $\overline{x}_{1 + 2}\big(p^\ast, \Delta^\ast\big)$ to decrease.  Since a large coverage factor allows the {\color{red} WSP} to charge a high access price $\Delta^\ast$ for Technology 2, {\color{red} WSP} revenue increases, offseting the decrease in savings from offloading less traffic.  We note, however, that this threshold $\eta$ value is largest for the dense city at about 80\%, and smallest for the small city at about 64\%.  This observation is consistent with the dense city's larger marginal savings from offloading.
 
In Fig. \ref{fig:costoffload_sparse}, we show the adoption levels $\overline{x}_1^\ast$ and $\overline{x}_{1 + 2}^\ast$ {\color{red} for the sparse city at the WSP}'s {\color{blue} profit-maximizing} operating point for a range of {\color{red} offloading benefits $c_{\rm WF}$. Results for the small and dense cities are similar.} We show the optimal coverage factors in Fig. \ref{fig:coverage_offload}; as we might expect {\color{red} from Fig. \ref{fig:etaall_opt}}, the optimal {\color{red} value of $\eta$} increases with {\color{red} $c_{\rm WF}$} and eventually becomes 1. 

\begin{figure}[t]
 \centering
 \includegraphics[width = 0.4\textwidth]{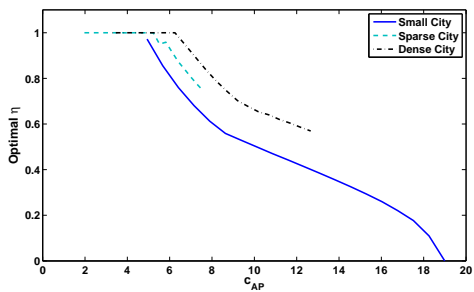}
 \vspace{-0.1in}
 \caption{Optimal coverage factors $\eta^\ast$ for the simulations in Fig. \ref{fig:costdeploy}.}
 \label{fig:coverage_deploy}
 \vspace{-0.2in}
 \end{figure}

We next vary the cost of deployment $c_{\rm AP}$, keeping $c_{\rm WF}$ constant.  Figure \ref{fig:costdeploy} shows the resulting adoption levels, with the optimal coverage factors $\eta^\ast$ shown in Fig. \ref{fig:coverage_deploy}. {\color{blue} For all cities, as the marginal cost of deployment $c_{\rm AP}$ increases, $\eta^\ast$ and the adoption $\overline{x}_1^\ast$ of Technology 1 decrease.  The decrease in $\eta^\ast$} is nonlinear: {\color{blue} initially, $\eta^\ast = 1$ as the cost of deployment is outweighed by the savings from offloading $c_{\rm WF}\eta\overline{x}_{1 + 2}^\ast$. The initial decrease from $\eta^\ast = 1$, however, is relatively rapid, as the decrease in $c_{\rm WF}\eta\overline{x}_{1 + 2}^\ast$ due to decreasing $\eta$ is muted by an increase in adoption of the bundled technologies $\overline{x}_{1 + 2}$.} For any fixed $c_{\rm AP}$, the optimal coverage factor is largest for the dense city and smallest for the small city; this reflects the ordering of the marginal savings from offloading $c_{\rm WF}$, which are largest for the dense city.

For the small city, the immediate decrease in the optimal coverage $\eta^\ast$ induces behavior similar to that of Fig. \ref{fig:maxrev_eta}: as $\eta^\ast$ decreases (Fig. \ref{fig:coverage_deploy}), {\color{red} adoption of the bundle first increases} and then decreases as Technology 2 experiences greter congestion {\color{blue} and less coverage}. The same effect is observed for the sparse and dense cities, without the final decrease in $\overline{x}_{1 + 2}^\ast$ for large $c_{\rm AP}$ (small $\eta^\ast$).  Thus, in cities with denser populations, a decrease in coverage due to higher costs may in fact increase adoption of the bundled technologies.

\section{Conclusion}\label{sec:conclusion}

In this paper, we develop a model of user adoption for base and supplemental wireless network technologies that accounts for heterogeneity in users' technology valuations, congestion effects, and pricing decisions.  We show that user adoption converges to a unique, stable equilibrium point, and derive analytical conditions under which non-intuitive adoption behaviors occur.  We then show that these may persist when {\color{red} WSP}s maximize either their revenue or profit.  {\color{red} Our profit model} incorporates both the savings from offloading and deployment costs of a WiFi network. % We find that the population density of the {\color{red} WSP}'s market can significantly affect the equilibrium adoption behavior.

Though we use empirical data to realistically study {\color{red} WSP} savings from offloading traffic onto a supplementary network, our parameters can only approximate true market structures.  Similarly, our user adoption model makes approximating assumptions, one of which is that users' technology valuations are uniformly distributed.  {\color{red} We therefore} show that many of the reported qualitative adoption behaviors are also observed for non-uniform distributions and nonlinear throughput functions (cf. {\color{red} Section \ref{sec:nonuniform_theta} and Appendix \ref{sec:throughput}}). Our framework thus enables researchers to gain additional insights into the role of supplementary technologies in alleviating network congestion and the resulting economic implications.

{\color{red} Broadly speaking, our work fits into the framework of ``Smart Data Pricing'' (SDP), which studies new ways of pricing mobile data with the aim of making mobile data networks more operationally efficient \cite{sen2013survey}. While other SDP works have shown that users change their behavior in response to pricing policies such as time-dependent pricing \cite{ha2012tube}, the framework introduced in this work can be used to investigate the effects of such pricing schemes on technology adoption levels when heterogeneous networks are present. Though such investigations are beyond the scope of this work, we anticipate that these smarter pricing policies can significantly improve WSP revenue and can even be used to steer the adoption levels to operationally {\color{blue} favorable} targets.} % We expect that further exploration and validation of this proposed analytical framework will yield additional insights into the benefits of congestion alleviation through offloading on supplementary technologies.

%Our work thus represents an initial investigation into {\color{red} WSP}s' profit maximization under user adoption of base and supplementary technologies.  %As such deployments increase, we expect that further analysis will yield additional insights

\bibliographystyle{IEEEtran}
\bibliography{Citations}

\appendices

\section{Proofs}\label{sec:proofs}

\subsection{Proposition \ref{prop:thetas}}

First we suppose that $\theta_{1,0} < \theta_{1 + 2, 0}$.  Then using (\ref{eq:theta3GWF}) and (\ref{eq:thetaWF}), the inequality $\theta_{1 + 2, 0} < \theta_{1,1 + 2}$ is equivalent to
 \begin{align}
 p\left(q_1 - q_2\right) > &q_1T_2\left(\eta x_{1 + 2}\right) - \eta^{-1}\Delta q_1 \nonumber \\ &- q_2T_1\left(x_1 + (1 - \eta) x_{1 + 2}\right). \label{eq:theta1}
 \end{align}
 But expanding the inequality $\theta_{1,0} < \theta_{1 + 2, 0}$ using (\ref{eq:theta3G}) and (\ref{eq:theta3GWF}), we have the exact same inequality.  On the other hand, if $\theta_{1 + 2, 0} < \theta_{1,0}$, then we switch the sign of (\ref{eq:theta1}) to show that $\theta_{1 + 2, 0} < \theta_{1,0}$ implies $\theta_{1,1 + 2} < \theta_{1 + 2, 0}$.
 \IEEEQED
 
\subsection{Proposition \ref{prop:stable}}

In regions a, f and g, it is simple to show that the equilibria are stable: letting $x = (x_1, x_{1 + 2})$, we denote (\ref{eq:xdyn}) as $\dot{x} = f(x)$ so that
\begin{equation*}
\frac{\partial f}{\partial x} = \begin{bmatrix} -1 & 0 \\ 0 & -1\end{bmatrix}
\end{equation*}
at each equilibrium in regions a, f and g.  Since this matrix is clearly negative definite, the equilibria are asymptotically stable.

In region b, the Jacobian of the dynamics (\ref{eq:xdyn}) is
\begin{equation*}
\begin{bmatrix} \frac{\gamma_1}{q_1 - q_2} -1 & \frac{-\eta\gamma_2 + (1 - \eta)\gamma_1}{q_1 - q_2} \\ \frac{-\gamma_1}{q_1 - q_2} & \frac{\eta\gamma_2 - (1 - \eta)\gamma_1}{q_1 - q_2} -1 \end{bmatrix}
\end{equation*}
The eigenvalues of this matrix are $\lambda = \eta\frac{\gamma_1 + \gamma_2}{q_1 - q_2} - 1$, $\lambda = -1$.  Since we assume $q_1 < q_2$, both eigenvalues are real and negative and the equilibrium in region b is stable.  Similarly, in region c the Jacobian has the form
\begin{equation*}
\begin{bmatrix}
\frac{\gamma_1}{q_1 - q_2} - \frac{\gamma_1}{q_1} - 1 & \frac{-\eta\gamma_2 + (1 - \eta)\gamma_1}{q_1 - q_2} - \frac{(1 - \eta)\gamma_1}{q_1} \\ \frac{-\gamma_1}{q_1 - q_2} & \frac{\eta\gamma_2 - (1 - \eta)\gamma_1}{q_1 - q_2} - 1
\end{bmatrix},
\end{equation*}
which has eigenvalues $\lambda$ satisfying
\begin{align*}
2\left(\lambda + 1\right) &= \frac{\eta\left(\gamma_2 + \gamma_1\right)}{q_1 - q_2} - \frac{\gamma_1}{q_1} \pm \\
&\sqrt{\left(\frac{\eta\left(\gamma_2 + \gamma_1\right)}{q_1 - q_2} - \frac{\gamma_1}{q_1}\right)^2 + \frac{4\eta\gamma_1\gamma_2}{q_1\left(q_1 - q_2\right)}}.
\end{align*}
Upon simplifying, we find that $\lambda = \eta\frac{\gamma_2 + \gamma_1}{q_1 - q_2}$, $-\frac{\gamma_1}{q_1}$.  As both of these are real and negative, we see that the equilibrium in region c is asymptotically stable.

Finally, in regions d and e we have the triangular Jacobian matrices
\begin{align*}
&\begin{bmatrix} \frac{-\gamma_1}{q_1} - 1 & \frac{-(1 - \eta)\gamma_1}{q_1} \\ 0 & -1 \end{bmatrix}, \\
&\begin{bmatrix} -1 & 0 \\ \frac{-(1 - \eta)\gamma_1}{(1 - \eta)q_1 + \eta q_2} & \frac{-(1  - \eta)^2\gamma_1 - \eta^2\gamma_2}{(1 - \eta)q_1 + \eta q_2} - 1 \end{bmatrix}.
\end{align*}
Since all diagonal entries in these matrices are negative, the equilibria in both regions are asymptotically stable.
\IEEEQED

% \subsection{Corollary \ref{cor:stable}}
%Each region is convex, and each point on the trajectory of $x_1(t)$ and $x_{1 + 2}(t)$ can be written as
%\begin{equation*}
%x(t) = x_0e^{-(t - t_0)} + \left(1 - e^{-(t - t_0)}\right)x^\star,
%\end{equation*}
%where $x^\star$ is the equilibrium point and $x_0$ the initial point.  Thus, $x(t)$ is a convex combination of the initial and equilibrium points.  The trajectory therefore lies within its region of origin, and therefore converges to the equilibrium point in that region.
%\IEEEQED

\subsection{Proposition \ref{prop:period}}
The result follows from Bendixson's criterion.  We can use the Jacobian expressions in the proof of Prop. \ref{prop:stable} to show that the divergence $\partial f_1/\partial x_1 + \partial f_2/\partial x_{1 + 2}$ of the dynamical equations (\ref{eq:xdyn}) is negative in each region.  Then Bendixson's criterion tells us that no periodic orbits can exist.\footnote{Bendixson's criterion relies on Green's theorem for the vector field $f$, which is typically assumed to be continuously differentiable.  In our case, $f$ is merely piecewise-linear and continuous.  However, since we are working in $\mathbb{R}^2$ and $f$ is continuously differentiable almost everywhere, the proof of Green's theorem is still valid.} % http://www.math.psu.edu/roe/230H/slides_14nov.pdf
Since no periodic orbits exist, each trajectory must converge to an equilibrium point. From Prop. \ref{prop:stable}, this equilibrium must be stable.
\IEEEQED

\subsection{Theorem \ref{thm:unique}}

The stability of the equilibrium follows from Prop. \ref{prop:stable}, while existence of at least one such equilibrium follows from the boundedness of the $x_i$ and the non-existence of a periodic orbit (Prop. \ref{prop:period}).  Thus, it remains to show that at most one equilibrium point can exist.

Suppose that two stable equilibrium points exist, and consider a bounded neighborhood $S$ of $\left\{x_1\geq 0, x_{1 + 2}\geq 0, x_1 + x_{1 + 2}\leq 1\right\}$.  We suppose that the dynamical equations (\ref{eq:xdyn}) are continuously extended to all of $S$ for the purposes of the proof, and that no new equilibria are created.  If we consider all trajectories lying in $S$, the regions of attraction for both equilibria are (disjoint) open, connected, and invariant sets, whose boundaries are formed by trajectories \cite{khalil}.  However, since no periodic orbits exist, the boundaries must be the boundary of $S$ itself.  Then since $S$ is connected, it cannot be a disjoint union of open sets, and there exists at least one point in $S$ that is in neither region of attraction.  But this is a contradiction, as the trajectory starting from this point must approach a compact limit set.
\IEEEQED

\subsection{Result \ref{prop:x2eta}}

We first note from Table \ref{tab:equilibria} that in regions a, b, f, and g of Table \ref{tab:regions}, the total adoption $\overline{x}_1 + \overline{x}_{1 + 2}$ remains constant at 1, i.e., full adoption. We may then differentiate the equilibrium expressions for $\overline{x}_1 + \overline{x}_{1 + 2}$ and $\overline{x}_{1 + 2}$ in region c and use the constraints in Table \ref{tab:equilibria} to show that these quantities are increasing in $\eta$. The remaining region is then region e, where $\overline{x}_1 = 0$ (no users adopt the base technology), $\overline{x}_1 + \overline{x}_{1 + 2} < 1$ (some users adopt neither technology), and $\overline{x}_{1 + 2} > 0$ (some users adopt the bundled technologies).  Differentiating Table \ref{tab:equilibria}'s equilibrium expression for $\overline{x}_{1 + 2}$ in this region, we find the condition (\ref{eq:eWFeta}).
\IEEEQED

\subsection{Result \ref{prop:p3G}}

We first note from Table \ref{tab:equilibria} that in regions a, b, e, f, and g in Table \ref{tab:regions}, adoption of the base technology ($\overline{x}_1$) does not depend on the price $p$. In region d, we differentiate Table \ref{tab:equilibria}'s expression for $\overline{x}_1$ to find that it is decreasing in $p$. Finally, we consider region c, where $\overline{x}_1 > 0$ (some users adopt the base technology), $\overline{x}_{1 + 2} > 0$ (some users adopt the bundled technologies), and $\overline{x}_1 + \overline{x}_{1 + 2} < 1$ (some users adopt neither technology).   We differentiate the equilibrium expression for $\overline{x}_1$ to obtain the condition (\ref{eq:p3G}).
\IEEEQED

\subsection{Proposition \ref{prop:rev_partial}}

By inspection, the optimal revenue in regions a, f, and g is non-positive in Table \ref{tab:revenue}. Thus, since full adoption only occurs in regions a, b, f, or g and the bundled technologies have positive adoption in regions c and e, for the first part of the proposition it suffices to show that the revenue in regions b and d does not exceed that in regions c or e.

{\bf Region b:} We first consider the case $2(1 - \eta)\gamma_1 - 2\eta\gamma_2 > q_2 - q_1$.  By inspection, in this case region b has negative revenue.  Since regions c, d and e have positive revenue, this cannot be optimal.

We now suppose that $2(1 - \eta)\gamma_1 - 2\eta\gamma_2 \leq q_2 - q_1$ and $\eta\gamma_2q_1< (1 - \eta)\gamma_1q_2$.  We show that the maximum revenue in region b is less than the maximum revenue in region e.  It suffices, from Table \ref{tab:revenue}, to show that
\begin{align*}
&\left(\eta\left(q_2 - q_1\right)^2 + 4(1 - \eta)\gamma_1\left(q_1 - q_2\right)\right) \\
&\times \left((1 - \eta)q_1 + \eta q_2 + (1 - \eta)^2 \gamma_1 + \eta^2\gamma_2\right) \\
&\leq \left((1 - \eta)q_1 + \eta q_2\right)^2\left(\eta\left(\gamma_1 + \gamma_2\right) + q_2 - q_1\right).
\end{align*}
Simplifying and neglecting some terms, we find the sufficient condition
\begin{align*}
&\eta \gamma_1q_1^2 + \eta\gamma_2 q_1^2 + q_1^2\left(q_2 - q_1\right) + 2\eta^2\gamma_1q_1\left(q_2 - q_1\right) \\
&+ 2\eta^2\gamma_2q_2\left(q_2 - q_1\right) + \eta q_1\left(q_2 - q_1\right)^2 + \eta^2\gamma_1\left(q_2 - q_1\right)^2 \\
&+ 4\eta(1 - \eta)\left(4 - 1 + \eta\right)\gamma_1\left(q_2 - q_1\right)^2 \geq 0,
\end{align*}
which is clearly true, since each term in the sum is nonnegative.

Next, we suppose that $\eta\gamma_2q_1\geq (1 - \eta)\gamma_1q_2$ and show that the revenue at region c's revenue-maximizing point is larger than the maximum revenue in region b.  From Table \ref{tab:revenue}, it suffices to show that
\begin{align*}
\frac{\frac{\eta}{4}\left(q_1 - q_2\right)^2}{\eta\left(\gamma_1 + \gamma_2\right) + q_2 - q_1} \leq &\frac{q_1^2\eta\gamma_2 + q_2^2\eta\gamma_1}{4A} \\
&+ \frac{q_1^2\left(q_2 - q_1\right)}{4A} \\
&+ \frac{\eta q_1\left(q_1 - q_2\right)^2}{4A}
\end{align*}
for any given $\eta$, where
\begin{align*}
A = &\gamma_1q_2 + \eta\gamma_1\gamma_2 \\ &+ q_1\left(q_2 - q_1 + \eta\gamma_2 - (1 - \eta)\gamma_1\right).
\end{align*}
Multiplying out the fractions, we find the sufficient condition
\begin{align*}
&\left(\eta\gamma_1q_2 + \eta^2\gamma_1\gamma_2\right)\left(q_1 - q_2\right)^2 \\
&- \eta(1 - \eta)q_1\gamma_1\left(q_1 - q_2\right)^2 \\
&\leq
\left(\eta\gamma_1 + \eta\gamma_2 + q_2 - q_1\right)\left(\eta\gamma_2q_1^2 + \eta\gamma_1q_2^2\right).
\end{align*}
We now expand the left side of this inequality; it is
\begin{align*}
\leq &-\eta\gamma_1q_1q_2^2 + \eta\gamma_1q_2^3 \\ &+ \eta^2\gamma_1\gamma_2q_1^2 - 2\eta^2\gamma_1\gamma_2q_1q_2 + \eta^2\gamma_1\gamma_2q_2^2 \\
= &\eta\gamma_1q_2^2\left(q_2 - q_1\right) + \eta^2\gamma_1\gamma_2q_1^2 \\ &- 2\eta^2\gamma_1\gamma_2q_1q_2 + \eta^2\gamma_1\gamma_2q_2^2.
\end{align*}
We now obtain
\begin{align*}
-2\eta^2\gamma_1\gamma_2q_1q_2 \leq 
&\eta^2\gamma_1^2q_2^2 + \eta^2\gamma_2^2q_1^2 \\ &+ \eta\gamma_1q_2^2\left(q_2 - q_1\right).
\end{align*}
Since the left-hand side is clearly negative, while the right-hand side is positive, we have the desired result.

{\bf Region d:} We first show that the revenue in region c is increasing in $\eta$.  By inspection, the revenue expressions in regions c, d, and e are equal at $\eta = 0$ and the optimal revenue in region d is independent of $\eta$, which will complete the proof.  We calculate that
\begin{align*}
\frac{dR_c}{d\eta} = &\frac{\gamma_1^2q_2^2\left(q_2 - q_1\right) + 2\gamma_1q_1q_2\left(q_2 - q_1\right)^2 }{4X^2} \\
&+ \frac{q_1^2\left(q_2 - q_1\right)^2 + (1 - \eta)\gamma_1q_1^3\left(q_2 - q_1\right)}{4X^2},
\end{align*}
where
\begin{align*}
X = &\gamma_1q_2 + \eta\gamma_1\gamma_2 \\
&+ q_1\left(q_2 - q_1 + \eta\gamma_2 - (1 - \eta)\gamma_1\right).
\end{align*}
Thus, $dR_c/d\eta \geq 0$ for all values of $\eta\in [0,1]$. The revenue in region c for $\eta = 1$ then exceeds that in region d. We note that since $\eta \gamma_2 q_1 \geq (1 - \eta)\gamma_1 q_2$ at $\eta = 1$, the expression in Table \ref{tab:revenue} for region c's maximum revenue is valid.

We next use the optimal revenue expressions in Table \ref{tab:revenue} to show that for fixed $\eta$, the revenue in region c exceeds that in region e if and only if
\begin{equation*}
\left((1 - \eta)\gamma_1q_2 - \eta q_1\gamma_2\right)^2\geq 0.
\end{equation*}
Since this expression is always nonnegative and region c's revenue is increasing in $\eta$, we see that the maximum revenue in region c occurs at $\eta = 1$, which exceeds that in region e. The maximum revenue thus occurs in region c, with $\eta = 1$.
\IEEEQED

%% From Appendix C
%\begin{figure}
%\centering
%\includegraphics[width = 0.45\textwidth]{Figures/RegionDEC_Beta_a1b3.eps}
%\vspace{-0.15in}
%\caption{Equilibrium adoption levels as the coverage factor $\eta$ increases, $\overline{x}_1 = 0$ for large $\eta$. User heterogeneity $\theta$ follows a $\beta$ distribution with parameters $(\alpha,\beta) = (1,3)$ (cf. Fig. \ref{fig:betadist}).  System parameters are $q_1 = 100$, $q_2 = 300$, $\gamma_1 = 50$, $\gamma_2 = 100$, $p = 40$, $\Delta = 10$.}
%\vspace{-0.15in}
%\label{fig:behavior_x10}
%\end{figure}
%
%\begin{figure}
%\centering
%\includegraphics[width = 0.45\textwidth]{Figures/BetaDist.eps}
%\vspace{-0.15in}
%\caption{Different $\beta$ distributions used for user heterogeneity $(\theta)$ in Fig. \ref{fig:behavior_nonuniform}.}
%\vspace{-0.15in}
%\label{fig:betadist}
%\end{figure}

\subsection{Proposition \ref{prop:revenuemax}}

%For $\eta\gamma_2q_1\geq (1 - \eta)\gamma_1q_2$, we show that the revenue in region e cannot exceed that in region c.  Under these conditions, the derivative with respect to $\eta$ of the maximum revenue in region e is nonnegative.  We then take the derivative of the revenue in region c with respect to $\eta$, and show that it is nonnegative for $\eta \in [0,1]$.  Since revenue in region c equals that in region d at $\eta = 0$ and the revenue in region d is not dependent on $\eta$, the proposition follows.

From the proof of Prop. \ref{prop:rev_partial}, we see that for fixed $\eta$, the maximum revenue occurs in region c if and only if
\begin{equation}
\eta q_1\gamma_2 > (1 - \eta)\gamma_1q_2,
\label{eq:covere}
\end{equation}
i.e., if (\ref{eq:covere}) holds then at the revenue-maximizing prices, the equilibrium adoption levels lie in region c. If (\ref{eq:covere}) does not hold, then the maximum revenue lies in either region d or region e. But we note that the revenue in regions d and e is the same at $\eta = 0$, and the revenue in region d does not depend on $\eta$. In contrast, in region e the revenue increases with $\eta$ if (\ref{eq:covere}) does not hold: we take the derivative of revenue in region e with respect to $\eta$ to find that it equals
\begin{equation*}
C\left[\eta\left(q_2 - q_1\right)^2 - 2\eta\gamma_1 q_2 - 2\eta\gamma_2 q_1 + 2\gamma_1q_2 + q_1\left(q_2 - q_1\right)\right],
\end{equation*}
where
\begin{equation*}
C = \frac{\left((1 - \eta)q_1 + \eta q_2\right)}{4\left((1 - \eta)q_1 + \eta q_2 + (1 - \eta)^2 \gamma_1 + \eta^2\gamma_2\right)^2}.
\end{equation*}
Since $\eta\gamma_2 q_1 \leq (1 - \eta)\gamma_1 q_2$, we see that this quantity is nonnegative, and thus that the revenue increases as $\eta$ increases.  Revenue in region e thus exceeds that in region d for any value of $\eta$ not satisfying (\ref{eq:covere}). No users adopt the base technology in region e (Table \ref{tab:equilibria}), which completes the proof.
%
%{\bf Region e:} We multiply the optimal revenue expressions in Table \ref{tab:revenue} and simplify terms to find that the revenue in region c exceeds that in region e if and only if
%%\begin{align*}
%%&-\eta^3\gamma_2q_1^3 - \eta(1 - \eta)\gamma_1q_1q_2^2 + \eta(1 - \eta)q_1^3\left(q_2 - q_1\right) \\ 
%%&+ \eta(1 - \eta)q_1^2\left(q_2 - q_1\right)^2 + \left(-\eta + 2\eta^2\right)q_1^2 q_2\left(q_2 - q_1\right) \\
%%&- \eta^2 q_1^2 q_2\left(q_2 - q_1\right) + \eta(1 - \eta)^2\gamma_1^2 q_2^2 \\
%%&- \eta(1 - \eta)^2\gamma_1 q_1^3 + \eta(1 - \eta)^2\gamma_1q_1\left(q_2 - q_1\right)^2 \\
%%&+ \eta^3\gamma_2^2 q_1^2 + 2\eta^3\gamma_2 q_1^2 q_2 + \eta^3\gamma_2 q_1\left(q_2 - q_1\right)^2 \\
%%&- 2\eta^2 (1 - \eta)\gamma_1\gamma_2 q_1q_2 + 2\eta(1 - \eta)^2\gamma_1q_1^2 q_2 \\
%%&- \eta^3 \gamma_2q_1q_2^2 + \eta^2(1 - \eta)\gamma_1q_1q_2^2 \geq 0.
%%\end{align*}
%\begin{equation*}
%\eta\left((1 - \eta)\gamma_1q_2 - \eta\gamma_2q_1\right)^2 \geq 0,
%\end{equation*}
%which clearly holds for all $\eta$.
%
%If $\eta\gamma_2q_1< (1 - \eta)\gamma_1q_2$, then the revenue-maximizing point lies in region e.  
\IEEEQED

\subsection{Result \ref{prop:optx}}

The ISP's revenue is maximized when the dynamics lie in region c if and only if $\eta\gamma_2q_1 \geq (1 - \eta)\gamma_1q_2$.  Thus, we can use Table \ref{tab:revenue}'s expressions for the optimal prices in region c to find the corresponding adoption levels in Table \ref{tab:equilibria}.  Differentiating with respect to $\eta$ shows that total adoption increases with $\eta$ in region c, yielding the proposition.
\IEEEQED

\section{Throughput Linearization}\label{sec:throughput}

In Section \ref{sec:usage}, we use linear models to represent the decrease in utility due to throughput degradation.  We justify this assumption here by analyzing the accuracy of a linear approximation to previously proposed throughput measures.

Prior works on technology adoption \cite{shetty2009economics} take congestion levels into account with a Markov chain analysis, assuming Poisson arrivals of rate $\lambda x$ and exponentially distributed session length with mean $\mu^{-1}$.  The expected throughput is then
\begin{equation}
-R_0(1 - \nu x)\frac{\log(1 - \nu x)}{\nu x},
\label{eq:throughputlog}
\end{equation}
where $R_0$ is the average time of service without interference or queueing, and {\color{red} we assume $\nu = \lambda/\mu < 1$}.  Thus, the \emph{throughput degradation} equals (\ref{eq:throughputlog}), less the maximum throughput.  We now use Taylor's remainder theorem to bound the error of a linear {\color{red} approximation to (\ref{eq:throughputlog}).  The second derivative of (\ref{eq:throughputlog}) is}
\begin{align*}
% R_0\left(\frac{-\gamma x -\log(1 - \gamma x)}{\gamma x^2}\right) = -R_0\sum_{i = 1}^\infty \frac{\gamma^ix^{i - 1}}{i + 1},
&R_0\left(\frac{2}{x^2} + \frac{\nu}{x} + \frac{\nu^2}{1 - \nu x} + \frac{2\log(1 - \nu x)}{\nu x^3}\right) = \\
&R_0\nu^2\sum_{n = 0}^\infty \frac{n + 1}{n + 3}\left(\nu x\right)^n,
\end{align*}
where the infinite sum uses the Taylor series for $\log (1 - \nu x)$ and the geometric series for $(1 - \nu x)^{-1}$.  Thus, approximating (\ref{eq:throughputlog}) at $x = 0.5$, we can bound the error by
\begin{align*}
&R_0\nu^2\max_{x\in(0,1)}\frac{\left(x - 0.5\right)^2}{2}\sum_{n = 0}^\infty \frac{n + 1}{n + 3}\left(\nu x\right)^n \\
&\leq \frac{R_0\nu^2}{8}\max_{x\in(0,1)}\sum_{n = 0}^\infty \frac{n + 1}{n + 3}\left(\nu x\right)^n.
\end{align*}
Approximating the sum as the geometric series of $\nu x$, we let $x = 1$ to obtain an upper bound of $R_0\nu^2/(8 - 8\nu)$. Numerically, this bound is in fact conservative; for instance, taking $R_0 = 1$ and $\nu = 0.5$ produces a maximum error of 0.013, as opposed to an analytical bound of 0.0625.

\begin{figure}
\centering
\includegraphics[width = 0.45\textwidth]{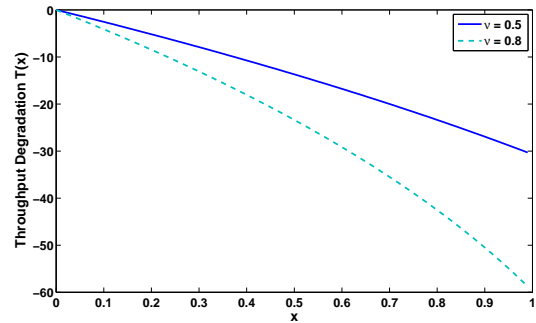}
\vspace{-0.15in}
\caption{Nonlinear throughput functions (\ref{eq:T_log}) used in Fig. \ref{fig:behavior_nonlinear}'s simulations.}
\vspace{-0.15in}
\label{fig:nonlinear}
\end{figure}

\begin{figure*}
\centering
\subfloat[Adoption with $\overline{x}_1 > 0$ for all $\eta$.]{\includegraphics[width = 0.35\textwidth]{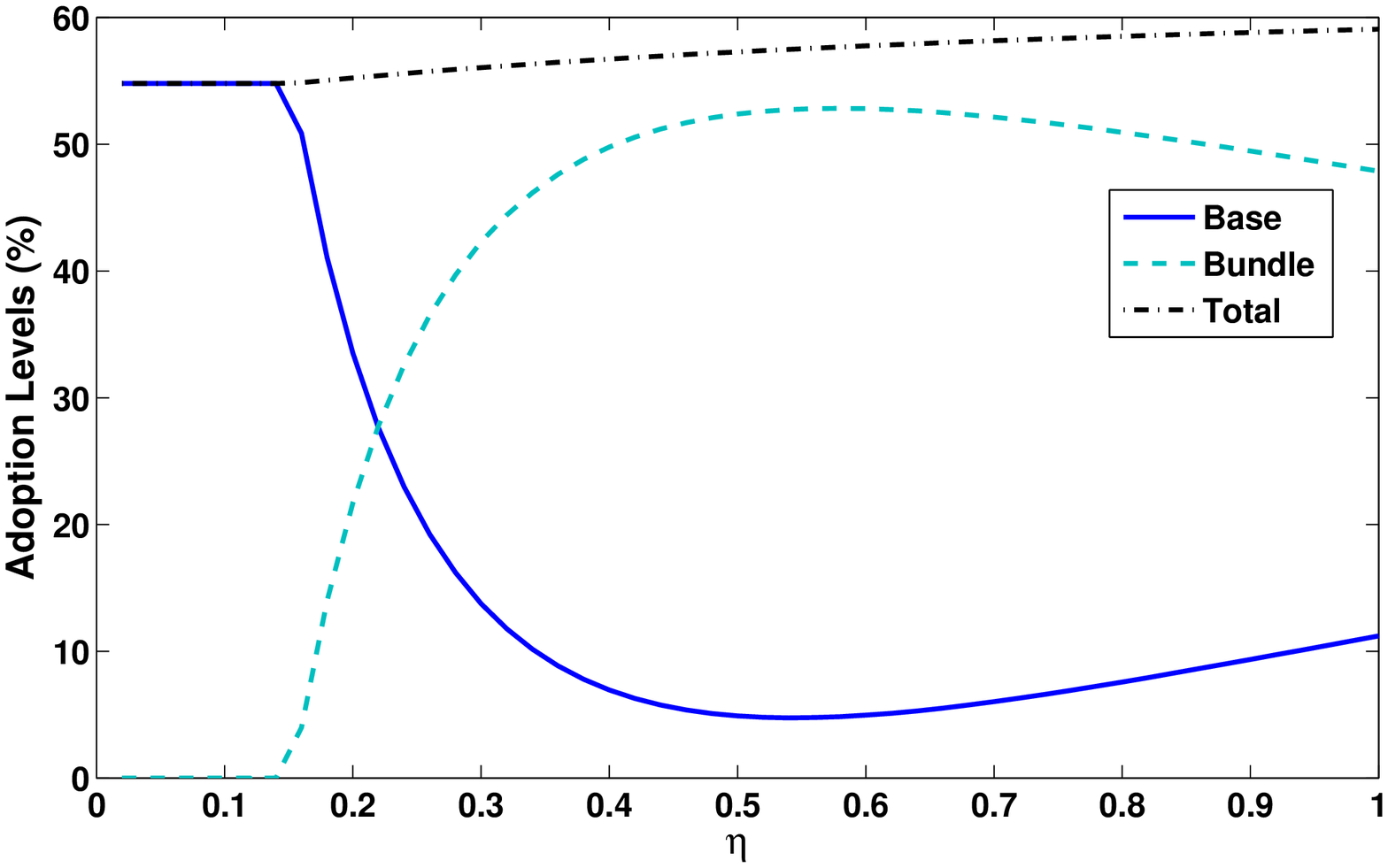}\label{fig:markovC}}\hspace{-0.035\textwidth}
\subfloat[Adoption with $\overline{x}_1 = 0$ for $\eta > 0.12$.]{\includegraphics[width = 0.35\textwidth]{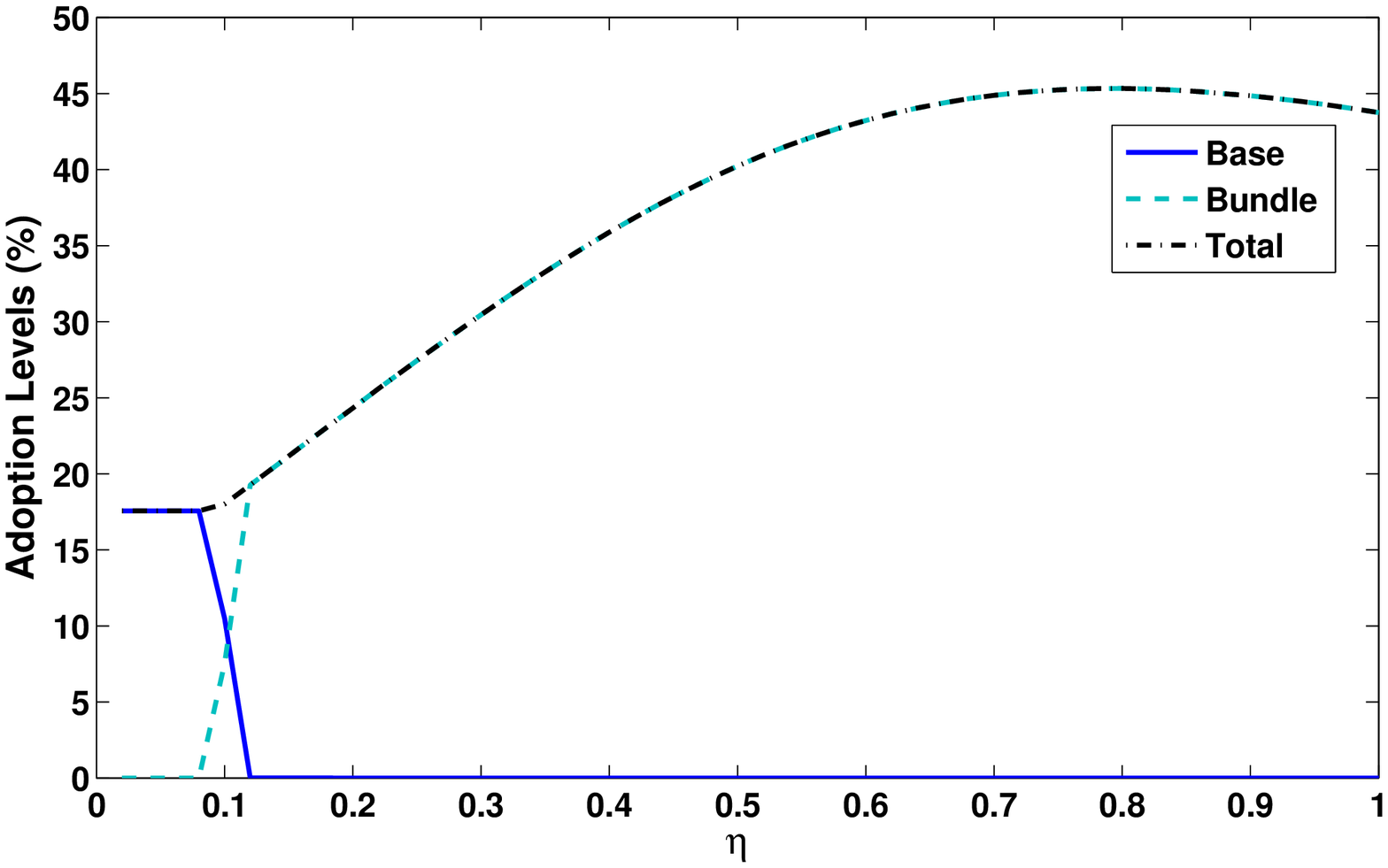}\label{fig:markovDEC}}\hspace{-0.035\textwidth}
\subfloat[Adoption with revenue-maximizing prices.]{\includegraphics[width = 0.35\textwidth]{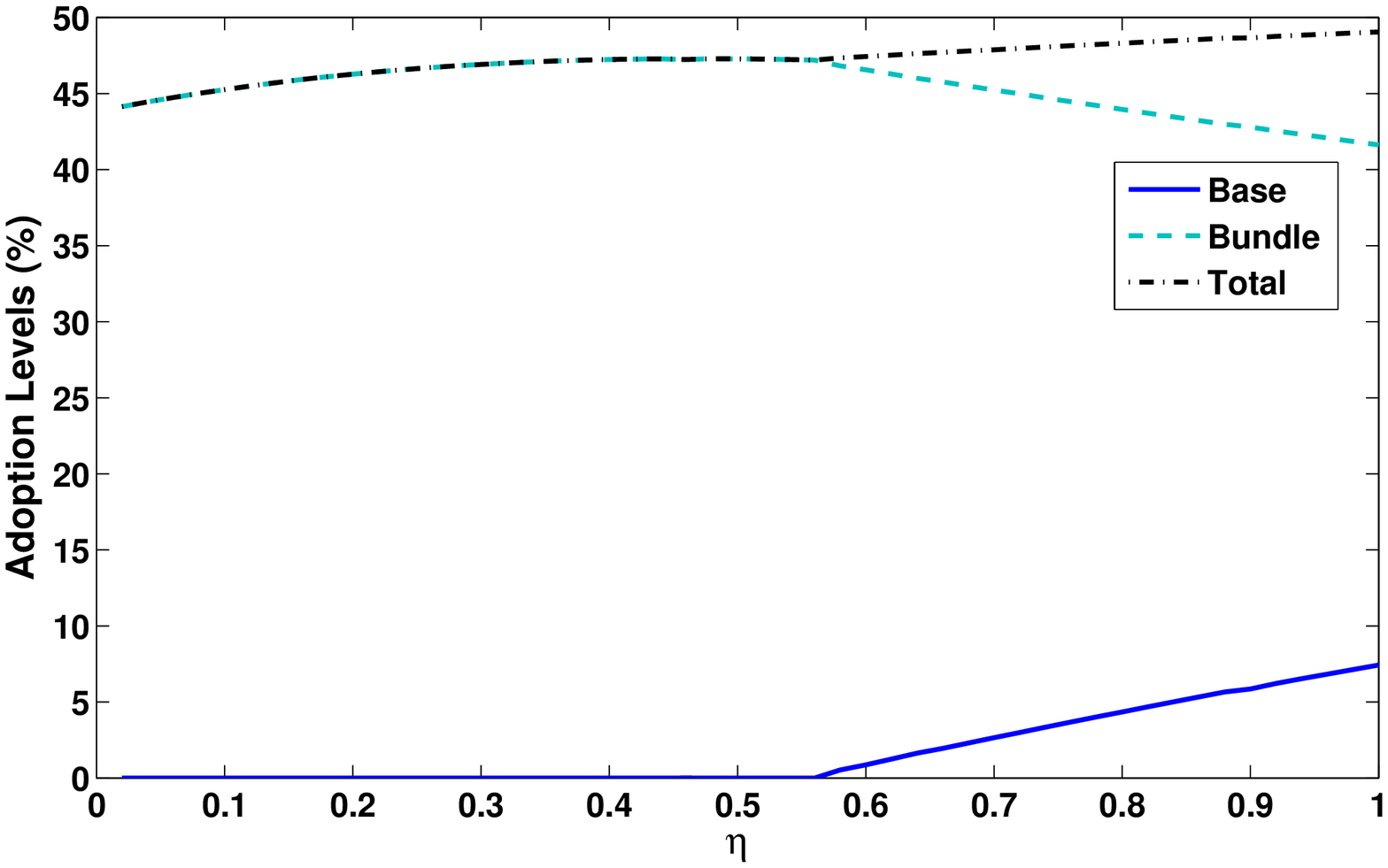}\label{fig:markovopt}}
\caption{Equilibrium adoption levels as the coverage factor $\eta$ varies for nonlinear throughput degradation functions $T_1$ and $T_2$ given by (\ref{eq:T_log}).  System parameters are (a) $q_1 = 50$, $q_2 = 80$, $R_1 = 10$, $R_2 = 50$, $\nu_1 = \nu_2 = 0.8$, $p= 20$, $\Delta = 5$; (b) $q_1 = 35$, $q_2 = 80$, $R_1 = 120$, $R_2 = 100$, $\nu_1 = \nu_2 = 0.8$, $p= 20$, $\Delta = 5$; and (c) $q_1 = 100$, $q_2 = 200$, $R_1 = 50$, $R_2 = 80$, $\nu_1 = \nu_2 = 0.5$, revenue-maximizing prices.}
\vspace{-0.2in}
\label{fig:behavior_nonlinear}
\end{figure*}

{\color{red} Having shown that linear functions can closely approximate proposed nonlinear throughput functions, we next investigate the dynamical behavior when the throughput degradation functions $T_1$ and $T_2$ are nonlinear. % We find that our results are qualitatively similar to those obtained with linear throughput functions.
We take the throughput functions to be as in (\ref{eq:throughputlog}), i.e.,}
% Given our results in Appendix \ref{sec:throughput} on the throughput approximation, we expect to observe similar adoption behaviors as for linear throughput functions, and we do in fact find that this is the case. We take the throughput functions to be those derived in Appendix \ref{sec:throughput}, i.e.,
\begin{equation}
T_i(x) = R_i\left(1 + (1 - \nu_i x)\frac{\log(1 - \nu_i x)}{\nu_i x}\right)
\label{eq:T_log}
\end{equation}
where $i = 1,2$ denotes the two technologies. Figure \ref{fig:nonlinear} shows the throughput functions (\ref{eq:T_log}) for $R_i = 100$ and $\nu_i = 0.5$, 0.8; a different $R_i$ coefficient merely scales these functions and does not affect their qualitative shape. As expected from Appendix \ref{sec:throughput}, these functions are close to linear, though their slope clearly increases for larger $x$ (i.e., more congestion).

Figure \ref{fig:behavior_nonlinear} shows a range of simulations with $\nu = 0.5$ (Figs. \ref{fig:markovC} and \ref{fig:markovDEC}) and $\nu = 0.8$ (Fig. \ref{fig:markovopt}) in (\ref{eq:T_log}). Figure \ref{fig:markovC} can be compared to Fig. \ref{fig:regionC}; in both figures, we observe a decrease in adoption of the bundled technologies and increase in adoption of the base technology for a large coverage factor $\eta$. Total adoption increases for all values of $\eta$ in both cases. Figures \ref{fig:regionDEC} and \ref{fig:markovDEC} show a similar decrease in adoption of the bundled technology for large $\eta$, but one in which adoption of the base technology is zero for larger values of $\eta$. Finally, Fig. \ref{fig:markovopt} shows that even when prices are chosen so as to maximize operator revenue, the adoption behavior with nonlinear throughput degradation functions mirrors that with linear throughput degradation, shown in Fig. \ref{fig:maxrev_eta}. In both cases, we observe that the adoption of the base technology is initially zero, with adoption of the bundled technologies $\overline{x}_{1 + 2}$ increasing with the coverage factor $\eta$. After a threshold value of $\eta$, however, adoption of the bundled technologies begins to decrease due to congestion on Technology 2, and adoption of Technology 1 correspondingly increases.

\section{Estimating Cost Parameters}\label{sec:costs}

{\it Savings from Offloading}:
We take the marginal cost of 3G traffic during the peak hour to be 1.0 \cent /MB in the small city, 1.9 \cent /MB in the sparse city, and 2.9 \cent /MB in the dense city; these values are based on AT\&T's and Verizon's data plan overage charges in the U.S.  We next estimate $C_{\rm WF}\eta$, the probability that a 3G and WiFi user has access to WiFi in the hour of peak 3G usage, multiplied by this peak 3G usage. From our trial data, we find that each user consumes on average 1200MB in each month, with 660MB occurring at peak hours of the day.  The probability of peak-hour WiFi access is 82\% of the overall access probability; thus, the probability of WiFi access during the peak hour is $0.82\eta$.  Each user then offloads ($0.82\eta$)(660MB) $= 541\eta$ MB at the peak hours over one month (i.e., $C_{\rm WF} = 541$).  {\color{red} Multiplying by the} \cent /MB marginal savings from offloading, we find that $c_{\rm WF} = 5.4$, 10.6, or 15.8 for the small, sparse, and dense cities respectively.

{\it Deployment Costs}:
To find the marginal cost of deployment $c_{\rm AP}$, we assume that each additional AP increases the coverage factor $\eta$ by a fixed amount $\Delta\eta$ and costs the {\color{red} WSP} a fixed amount $C_{\rm AP}$ per month. From \cite{wireless2020}, we estimate $C_{\rm AP}$ as a monthly operational cost of \$20, plus capital investment of \$1200 spread over 12 months, so that $C_{\rm AP}$ = \$120.  For simplicity, we interpret the WiFi access probability $\eta$ as the physical area covered by APs, e.g., uniform user mobility.  The cost of covering an area $\eta$ with APs is then $C_{\rm AP}\lceil \eta/\Delta\eta\rceil \approx \left(C_{\rm AP}/\Delta\eta\right)\eta$.  Normalizing by the user population, we find that
{\allowdisplaybreaks \begin{align*}
c_{\rm AP}\eta &= \frac{C_{\rm AP}\left({\rm Market\;area}\right)}{\left({\rm AP\;coverage\;area}\right)\left({\rm Market\;population}\right)}\eta \\
&= \frac{\$120}{\left({\rm AP\;coverage\;area}\right)\left({\rm Population\;density}\right)}\eta.
\end{align*}}
We use population densities of 2000, 5000, and 12000 people per square mile for the small, sparse, and dense cities respectively.  The AP coverage area is assumed to be 0.01 square miles, (a 130 meter radius), for the small city, 0.005 square miles for the sparse city, and 0.002 square miles for the dense city.  We then find $c_{\rm AP} = 6.2$, 4.9, and 11.5 for the small, sparse, and dense cities respectively.

{\color{red}

\section{Extension to Multiple WSPs}\label{sec:multiple}

We can extend our model to include multiple WSPs by adding additional adoption choices for users, e.g., adopting the base technology from another WSP. Conceptually, this change is simple: we simply formulate an additional utility function for each additional adoption choice, and then find the option that yields the highest utility for each value of $\theta(v)$. In practice, solving for the optimal choice for each user becomes complex due to the many possible orderings of the utility functions for different values of $\theta$ (i.e., extending Proposition \ref{prop:thetas} to multiple adoption choices). In this Appendix, we illustrate this complexity by considering two {\color{red} WSP}s: one offering both the base and bundled technologies, and the other offering only the base technology. Each user then has four adoption choices: adopt no technology, adopt the base technology 1 from {\color{red} WSP} 1, adopt the bundled technologies (1 + 2) from {\color{red} WSP} 1, or adopt the base technology 3 from {\color{red} WSP} 2.

We suppose that the utility of {\color{red} WSP} 1's base and bundled technologies is given by (\ref{eq:utility3g}) and (\ref{eq:utilitywf}) as in Section \ref{sec:usage}; the utility of {\color{red} WSP} 2's base technology is given analogously by
\begin{equation}
U_3 = \theta q_3 + T_3(x_3) - p_3
\end{equation}
where as in Section \ref{sec:usage} the intrinsic quality of the technology is captured in the variable $q_3$, $T_3$ represents the negative externality experienced due to {\color{blue} greater traffic on Technology 3, and $p_3$ is the access price}. The fraction of users adopting Technology 3 is denoted by $x_3(t)$.

We next solve for the utility-maximizing adoption choice as a function of $\theta$. As in Section \ref{sec:usage}, we solve for the threshold $\theta$ values at which the ordering of different adoption choices changes. The thresholds $\theta_{(1,0)}$, $\theta_{(1 + 2,0)}$, and $\theta_{(1 + 2,1)}$ are as in Section \ref{sec:usage}; the remaining thresholds are given by
\begin{align*}
\theta_{(3,0)} = &\frac{p_3 - T_3(x_3)}{q_3} \\
\theta_{(3,1)} = &\frac{T_1\left(x_1 + (1 - \eta)x_{1 + 2}\right) - T_3(x_3) - p + p_3}{q_3 - q_1} \\
\theta_{(1 + 2,3)} = &\frac{T_3(x_3) - (1 - \eta)T_1\left(x_1 + (1 - \eta)x_{1 + 2}\right)}{(1 - \eta)q_1 + \eta q_2 - q_3} \\
& \frac{- \eta T_2\left(\eta x_{1 + 2}\right) + p + \Delta - p_3}{(1 - \eta)q_1 + \eta q_2 - q_3}
\end{align*}
where $\theta_{(x,y)}$ again denotes the threshold value of $\theta$ above which adoption choice $x$ yields higher utility than choice $y$. To keep this meaning consistent, we suppose that $q_1 < q_3 < (1 - \eta)q_1 + \eta q_2$; if instead $q_3 < q_1$, the threshold $\theta_{(3,1)}$ should instead be denoted $\theta_{(1,3)}$, and if $q_3 > (1 - \eta)q_1 + \eta q_2$, the threshold $\theta_{(1 + 2,3)}$ would become $\theta_{(3, 1 + 2)}$.

We next derive the analogue of Proposition \ref{prop:thetas} by enumerating the possible orderings of the $\theta$ thresholds:
\begin{proposition}\label{prop:mult}
Suppose we are given the pairwise ordering of the following pairs of thresholds: (a) $\theta_{(1,0)}$ and $\theta_{(1 + 2,0)}$; (b) $\theta_{(1,0)}$ and $\theta_{(3,0)}$; (c) $\theta_{(3,0)}$ and $\theta_{(1 + 2,0)}$; (d) $\theta_{(3,1)}$ and $\theta_{(1 + 2,1)}$; (e) $\theta_{(3,0)}$ and $\theta_{(1 + 2,1)}$; (f) $\theta_{(3,1)}$ and $\theta_{(1 + 2,0)}$; and (g) $\theta_{(1 + 2,3)}$ and $\theta_{(1,0)}$. Then for any user with $\theta\in[0,1]$, we can determine the adoption choice yielding the highest utility for this user.
\end{proposition}
\begin{IEEEproof}
First, we show that the orderings (a--d) above imply the following:
\begin{enumerate}[(a)]
\item
$\theta_{(1,0)} < \theta_{(1 + 2,0)}$ if and only if $\theta_{(1 + 2,0)} < \theta_{(1 + 2,1)}$.

This is exactly the statement of Proposition \ref{prop:thetas}.
\item
$\theta_{(1,0)} < \theta_{(3,0)}$ if and only if $\theta_{(3,0)} < \theta_{(3,1)}$.

Suppose that $\theta_{(1,0)} < \theta_{(3,0)}$. Then upon cross-multiplying the denominators and simplifying, $\theta_{(1,0)} < \theta_{(3,0)}$ is equivalent to
\begin{equation*}
q_3 p + q_3\gamma_1\left(x_1 + (1 - \eta)x_{1 + 2}\right) < q_1p_3 + q_1\gamma_3x_3,
\end{equation*}
which is exactly the same expression obtained by simplifying the expression $\theta_{(3,0)} < \theta_{(3,1)}$. Thus, the two inequalities are equivalent.
\item
$\theta_{(3,0)} < \theta_{(1 + 2,0)}$ if and only if $\theta_{(1 + 2,0)} < \theta_{(1 + 2,3)}$.

Suppose that $\theta_{(3,0)} < \theta_{(1 + 2,0)}$. We can cross-multiply and simplify to find the equivalent inequality
\begin{align*}
p_3&\left((1 - \eta)q_1 + \eta q_2\right) + \gamma_3x_3\left((1 - \eta)q_1 + \eta q_2\right) < \\
&(1 - \eta)q_3\gamma_1\left(x_1 + (1 - \eta)x_{1 + 2}\right) \\
&+ q_3\eta^2 \gamma_2 x_{1 + 2} + pq_3 + \Delta q_3.
\end{align*}
We can obtain the same inequality by simplifying $\theta_{(1 + 2,0)} < \theta_{(1 + 2,3)}$. As in case (b), we thus obtain the proof in the other direction.
\item
$\theta_{(3,1)} < \theta_{(1 + 2,1)}$ if and only if $\theta_{(1 + 2,1)} < \theta_{(1 + 2,3)}$.

As in cases (b) and (c) above, we first cross-multiply and simplify $\theta_{(3,1)} < \theta_{(1 + 2,1)}$ to obtain
\begin{align*}
-\eta &q_2\gamma_1\left(x_1 + (1 - \eta)x_{1 + 2}\right) + \eta\left(\gamma_3x_3 - p + p_3\right)\left(q_2 - q_1\right) \\
&\left(\eta^2\gamma_2 x_{1 + 2} + \Delta\right)\left(q_3 - q_1\right) - \eta q_3\gamma_1\left(x_1 + (1 - \eta)x_{1 + 2}\right)
\end{align*}
Simplifying $\theta_{(1 + 2,1)} < \theta_{(1 + 2,3)}$ yields the same inequality, thus yielding the other direction of the proof.
\end{enumerate}
We next show that for any two thresholds $\theta_{(w,x)}$ and $\theta_{(y,z)}$ where $w,x,y,z\in\left\{0, 1, 1 + 2, 3\right\}$, the orderings (a--d) above allow us to determine whether $\theta_{(w,x)} < \theta_{(y,z)}$ or vice versa so long as the $w,x,y,z$ are not all distinct. We first note that at most two of $w,x,y,z$ can be equal, since $w \neq x$ and $y\neq z$ for all thresholds. We now divide the proof into three cases:
\begin{itemize}
\item
Suppose that $w = y$. Then without loss of generality, $x = 1$ and $z = 0$. If $w = y = 3$ or $1 + 2$, then orderings (b) and (a) respectively determine whether $\theta_{(w,x)} < \theta_{(y,z)}$ or $\theta_{(y,z)} < \theta_{(w,x)}$.
\item
Suppose that $w = z$. Then $w = 1$ or $w = 3$.

If $w = 1$, then $x = 0$ and $y = 3$ or $y = 1 + 2$. We must show that orderings (a-d) determine whether $\theta_{(1,0)} < \theta_{(y,1)}$ or $\theta_{(1,0)} > \theta_{(y,1)}$, which is determined by orderings (b) and (a) if $y = 3$ or $y = 1 + 2$ respectively.

If $w = 3$, then $y = 1 + 2$ and $x = 1$ or $0$. Orderings (c) and (d) respectively determine whether $\theta_{(3,x)} < \theta_{(1 + 2,3)}$ or $\theta_{(3,x)} > \theta_{(1 + 2,3)}$ if $x = 0$ or $x = 1$ respectively.
\item
Suppose that $x = z$. Then $x = 1$ or $x = 0$, and without loss of generality $w = 3$ and $y = 1 + 2$. Then whether $\theta_{(3,x)} < \theta_{(1 + 2,x)}$ or $\theta_{(3,x)} > \theta_{(1 + 2,x)}$ is determined by orderings (c) and (d) respectively if $x = 0$ or $x = 1$.
\end{itemize}
We can now conclude that given two thresholds $\theta_{(w,x)}$ and $\theta_{(y,z)}$, their relative order is determined by inequalities (a--d) unless $w,x,y,z$ are distinct. In that case, since there are only four possible values for $w,x,y,z$, $\theta_{(w,x)}$ and $\theta_{(y,z)}$ are ordered with respect to all other thresholds. Thus, we can order all six thresholds given inequalities (a--d) except in the case where no other threshold lies between $\theta_{(w,x)}$ and $\theta_{(y,z)}$,
%
% Finally, we show that if $w,x,y,z$ are all distinct and $\theta_{(w,x)}$ and $\theta_{(y,z)}$ have no other threshold between them, then the relative order of $\theta_{(w,x)}$ and $\theta_{(y,z)}$ does not change the adoption choice of any user.
in which case we must use orderings (e--g) to find the ordering of $\theta_{(w,x)}$ and $\theta_{(y,z)}$. Since any user's adoption choice is wholly determined by the orderings of the $\theta$ thresholds relative to the user's value of $\theta$, the proposition follows: the {\color{blue} orderings} (a--g) completely determine the ordering of the $\theta$ thresholds.
%
% Suppose that $w,x,y,z$ are all distinct and there is no other threshold between $\theta_{(w,x)}$ and $\theta_{(y,z)}$. Suppose that $\theta_{(w,x)} < \theta_{(y,z)}$. We will show that for any value of $\theta$, the utility-maximizing adoption choice does not change if we instead have $\theta_{(y,z)} < \theta_{(w,x)}$. If $\theta < \theta_{(w,x)} < \theta_{(y,z)}$ or $\theta > \theta_{(y,z)} > \theta_{(w,x)}$, then clearly this is the case. If $\theta_{(w,x)} < \theta < \theta_{(y,z)}$, then the user will choose $w$ over $x$ but $z$ over $y$; the user's adoption choice is then determined by $\theta_{(w,z)}$. If instead $\theta_{(y,z)} < \theta < \theta_{(w,x)}$, then the user will choose $y$ over $z$ and $x$ over $w$; the user's adoption choice is then determined by $\theta_{(y,x)}$. % If (w,z) > (y,z), then the user chooses z. 
\end{IEEEproof}
We next note that the seven orderings (a--g) cannot be chosen independently; for instance, if orderings (b) and (c) imply that $\theta_{(1,0)} < \theta_{(3,0)} < \theta_{(1 + 2,0)}$, then ordering (a) must be $\theta_{(1,0)} < \theta_{(1 + 2,0)}$. To enumerate all possible orderings, we introduce the ``sign'' of an ordering, which is $+$ if the ordering is the same as stated in Proposition \ref{prop:mult}, and $-$ if reversed. For instance, ordering (a) is $+$ if $\theta_{(1,0)} < \theta_{(1 + 2,0)}$.

We first note that orderings (a) and (b) are independent. Ordering (c) is determined by orderings (a) and (b) if these orderings have opposite sign, so orderings (a--c) have 6 possible orderings total. From the proof of Prop. \ref{prop:mult}, ordering (d) is similarly determined by orderings (a--c) if any pair of these have opposite signs. Thus, orderings (a--d) yield 8 possible types of orderings for the six $\theta$ thresholds.

Orderings (e--g) can also not be chosen independently of orderings (a--d); these orderings serve as ``tiebreakers'' in the case that two thresholds with non-overlapping components end up placed consecutively. By explicitly enumerating the 8 possibilities of orderings (a--d), we find that two orderings require one tiebreaker each, and two others require two tiebreakers each. We thus find a total of 16 orderings of the six $\theta$ thresholds, which are as follows:
\begin{enumerate}
\item
$\theta_{(1,0)} < \theta_{(3,0)} < \left(\theta_{(1 + 2,0)}, \theta_{(3,1)}\right) < \theta_{(1 + 2,1)} < \theta_{(1 + 2,3)}$
\item
$\theta_{(1,0)} < \theta_{(3,0)} < \theta_{(1 + 2,0)} < \theta_{(1 + 2,3)} < \theta_{(1 + 2,1)} < \theta_{(3,1)}$
\item
$\left(\theta_{(1,0)}, \theta_{(1 + 2,3)}\right) < \theta_{(1 + 2,0)} < \left(\theta_{(3,0)}, \theta_{(1 + 2,1)}\right) < \theta_{(3,1)}$
\item
$\theta_{(1 + 2,3)} < \theta_{(1 + 2,1)} < \left(\theta_{(3,1)}, \theta_{(1 + 2,0)}\right) < \theta_{(3,0)} < \theta_{(1,0)}$
\item
$\theta_{(3,1)} < \theta_{(1 + 2,1)} < \theta_{(1 + 2,3)} < \theta_{(1 + 2,0)} < \theta_{(3,0)} < \theta_{(1,0)}$
\item
$\theta_{(3,1)} < \left(\theta_{(3,0)}, \theta_{(1 + 2,1)}\right) < \theta_{(1 + 2,0)} < \left(\theta_{(1,0)}, \theta_{(1 + 2,3)}\right)$
\item
$\theta_{(1 + 2,3)} < \theta_{(1 + 2,1)} < \theta_{(1 + 2,0)} < \theta_{(1,0)} < \theta_{(3,0)} < \theta_{(3,1)}$
\item
$\theta_{(3,1)} < \theta_{(3,0)} < \theta_{(1,0)} < \theta_{(1 + 2,0)} < \theta_{(1 + 2,1)} < \theta_{(1 + 2,3)}$
\end{enumerate}

We use the notation $\left(\theta_{(w,x)}, \theta_{(y,z)}\right)$ to indicate that $\theta_{(w,x)} < \theta_{(y,z)}$ and $\theta_{(w,x)} > \theta_{(y,z)}$ are both valid pairwise orderings. Given these threshold orderings, one can then characterize different regions as in Table \ref{tab:regions} and solve for each region's adoption dynamics and equilibrium adoption point. It is then straightforward to solve for the equilibrium in each region and derive conditions under which each equilibrium point exists. One can then use such equilibria to examine, e.g., what proportion of users {\color{red} WSP} 1 will gain from {\color{red} WSP} 2 by introducing the supplementary technology.
}

\end{document}